\documentclass[noshowpacs,pra,aps,superscriptaddress,floatfix,twocolumn,nofootinbib]{revtex4-2} 
\usepackage[utf8]{inputenc}
\usepackage{lineno,mathtools,url}
\usepackage{amsmath}
\usepackage[normalem]{ulem}
\usepackage{amssymb,mathrsfs}
\usepackage{graphicx}
\usepackage{hyperref}
\hypersetup{
     colorlinks=true,
     linkcolor=blue,
     filecolor=blue,
     citecolor = orange,      
   }
\usepackage[dvipsnames]{xcolor}
\usepackage{dsfont}
\usepackage{mathbbol} 
\newcommand{\sss}{\scriptscriptstyle}

\begin{document}

\title{Violation of Bell inequalities in an analogue black hole}
\author{Giorgio Ciliberto}
\affiliation{Universit\'e Paris-Saclay, CNRS, LPTMS, 91405, Orsay, France}
\affiliation{Physikalisches Institut, Albert-Ludwigs-Universit\"at Freiburg,
Hermann-Herder-Stra{\ss}e 3, 79104 Freiburg, Germany}
\author{Stephanie Emig}
\affiliation{Physikalisches Institut, Albert-Ludwigs-Universit\"at Freiburg,
Hermann-Herder-Stra{\ss}e 3, 79104 Freiburg, Germany}
\author{Nicolas Pavloff}
\affiliation{Universit\'e Paris-Saclay, CNRS, LPTMS, 91405, Orsay, France}
\affiliation{Institut Universitaire de France}
\author{Mathieu Isoard}
\affiliation{Laboratoire Kastler Brossel, Sorbonne Universit\'e, CNRS, ENS-PSL Research University,
Coll\`ege de France, 4 place Jussieu, F-75252 Paris, France}
\begin{abstract}
  Signals of entanglement and nonlocality are quantitatively evaluated
  at zero and finite temperature in an analogue black hole realized in
  the flow of a quasi one-dimensional Bose-Einstein condensate. The
  violation of Lorentz invariance inherent to this analog system opens
  the prospect to observe 3-mode quantum correlations and we study the
  corresponding violation of bipartite and tripartite Bell
  inequalities. It is shown that the long wavelength modes of the
  system are maximally entangled, in the sense that they realize a
  superposition of continuous variable versions of
  Greenberger-Horne-Zeilinger states whose entanglement resists
  partial tracing.
\end{abstract}

\maketitle

\section{Introduction}\label{sec:intro}

The domain of analog gravity aims at providing laboratory models for
gaining insight on general relativity phenomena that cannot be
directly observed in the usual gravitational context or for which
there exists no complete theoretical framework. Two such phenomena are
black hole superradiance
\cite{Penrose1971,Zeldovich1971,Zeldovich1972} and Hawking radiation
\cite{Hawking1974,Unruh_1981}. In that context, the most successful
analogous experimental platforms have been surface gravity
\cite{Weinfurtner2011,Euve2016,Torres2017,Euve2020,Torres2020} and
acoustic \cite{Cromb2020} waves, nonlinear optical systems \cite{Philbin2008,Elazar2012,Drori2019,Braidotti2022}, cavity polaritons
\cite{Nguyen2015,Jacquet2020b,Falque2023} and Bose-Einstein
condendates (BEC) of atomic vapors
\cite{Lahav2010,De_nova_2019,Kolobov_2021}. Because of their low
temperature and coherence properties BECs appear particularly well
suited for demonstrating quantum features. 
We will not address here the question of the actual experimental demonstration of quantum entanglement in a BEC analogue black hole
(see, e.g., Ref. \cite{Isoard2021} for a recent discussion) but rather
take the theoretical analysis a little further by asking: which
general insight can we reach by studying quantum correlations of the
Hawking signal emitted by an analogue black hole? A natural approach
for such an investigation is a test of nonlocality via violation of
Bell inequality. The epistemological query of refutation of local
hidden variable theories has already received an unambiguous answer in
many contexts (see, e.g., Ref. \cite{Brunner2014} and references
therein) and important progresses have also been achieved in the field
of BEC matter waves
\cite{Schmied2016,Fadel2018,Kunkel2018,Colciaghi2023} we consider
here. Such a test is nonetheless a nontrivial extension of the scope
of analogue gravity and would constitute a primer for continuous
variables entanglement in a matter wave environment (see also Refs.
\cite{Piano2013,Lewis2015,Thomas2022} for related proposals).  In view
of future experimental studies it is relevant to quantitatively
evaluate in realistic configurations to what extent Bell inequalities
can be violated in BEC analogues. This is a natural question to ask,
all the more so as we argue in the following that in some (exotic)
limits the analog black hole we consider exactly realizes a
Einstein-Podolsky-Rosen (EPR) pair.  Furthermore, as will be shown,
the specifics of the system provide a natural testing ground for
genuine {\it tripartite} nonlocality. Our theoretical investigation of
the matter reveals an unexpected generic feature of black hole
analogues: in the long wave-length limit, the state of the system
realizes an infinite sum of degenerate Greenberger-Horne-Zeilinger
(GHZ) states. Interestingly, thanks to the continuous nature of its
degrees of freedom, and despite the clear GHZ nature of its long
wavelength modes, the analog system remains entangled after partial
tracing.

The paper is organized as follows. Section \ref{sec.model} presents the BEC analogue system we consider and the theoretical tools we employ for its study. Section \ref{sec.Bell2} is devoted to the study of measures of bipartite nonlocality and Section \ref{sec.Bell3} to tripartite nonlocality. We present our conclusions in Sec. \ref{sec.conclusion}. Technical aspects are summarized in the appendices. Appendices \ref{app.different} and \ref{app.sigma} are devoted to a brief presentation of previous results. 
The black hole analogue we consider can be modeled by an equivalent optical system 
 \cite{Isoard2021} whose relevance for the aspects we consider in the main text is discussed in Appendix \ref{app:analogue}. Appendix \ref{app.pseudo.spin} present elementary properties of the pseudo-spin we use in the main text. Appendices \ref{app.maxi2}, 
\ref{app.maxi3} and \ref{app.Maximization} treat specific technical aspects of the procedures used in Secs. \ref{sec.Bell2}  and \ref{sec.Bell3} for maximizing the violation of bipartite and tripartite Bell inequalities. Appendix \ref{app.averages} also concerns technical aspects, but these are of crucial importance for the computation of all the quantities we present and plot in the main text, such as the Clauser-Horne-Shimony-Holt (CHSH), Svetlichny and Mermin parameters. The last appendix (\ref{app.other}) presents results in analogue black hole configurations different from the one on which we focus in the main text.

\section{The model}\label{sec.model}

We consider a one-dimensional (1D) BEC described by a quantum field $ \hat\Psi(x,t)$ solution of
\begin{equation}\label{eq.model1}
    i\hbar \partial_t \hat\Psi = -\frac{\hbar^2}{2m}\partial_x^2\hat\Psi
    +[U(x) + g \, \hat{n} - \mu]\hat\Psi .
\end{equation}
In this equation $U(x)$ is an external potential whose precise form depends
on the black hole configuration considered, $\mu$ is the chemical potential, $\hat{n}(x,t)=\hat\Psi^\dagger\hat{\Psi}$ is the density operator and $g$ is the nonlinear constant describing the effective interaction within the system. We consider an effective repulsion, with $g>0$. In the so-called 1D mean-field regime \cite{Menotti2002} the quantum field can be written as
\begin{equation}\label{eq.model2}
    \hat\Psi(x,t)=\Phi(x)+\hat\psi(x,t),
\end{equation}
where $\Phi(x)$ describes the background flow while $\hat\psi$
accounts for small quantum fluctuations. Whereas the decomposition
\eqref{eq.model2} is legitimate in 3D, it has a finite range of
validity in the 1D configurations we consider; however, its conditions
of applicability are commonly met in standard experimental situations,
see, e.g., the discussion in \cite{Fabbri2018} which is summarized in Appendix \ref{app.different}. In decomposition
\eqref{eq.model2} the classical field $\Phi$ is a stationary solution
of the Gross-Pitaevskii equation, see Eq. \eqref{eq.model1bis}.  The
associated flow realizes an analogue black hole horizon if it is
subsonic in the asymptotic upstream region and supersonic in the
asymptotic downstream region. Several such flows have been considered
in the past. We will here focus on the ``waterfall configuration''
which is close to the experimental realization of
Refs. \cite{steinhauer_2016,De_nova_2019}. In order to assess the
generality of our results, we present in Appendix \ref{app.other} the
results obtained for two other configurations, we denote as ``delta
peak'' and ``flat profile''.  All these configurations are described in
Appendix \ref{app.different}.

\subsection{Propagation channels and quantum modes}

\begin{figure}
\centering
\includegraphics[width=\linewidth]{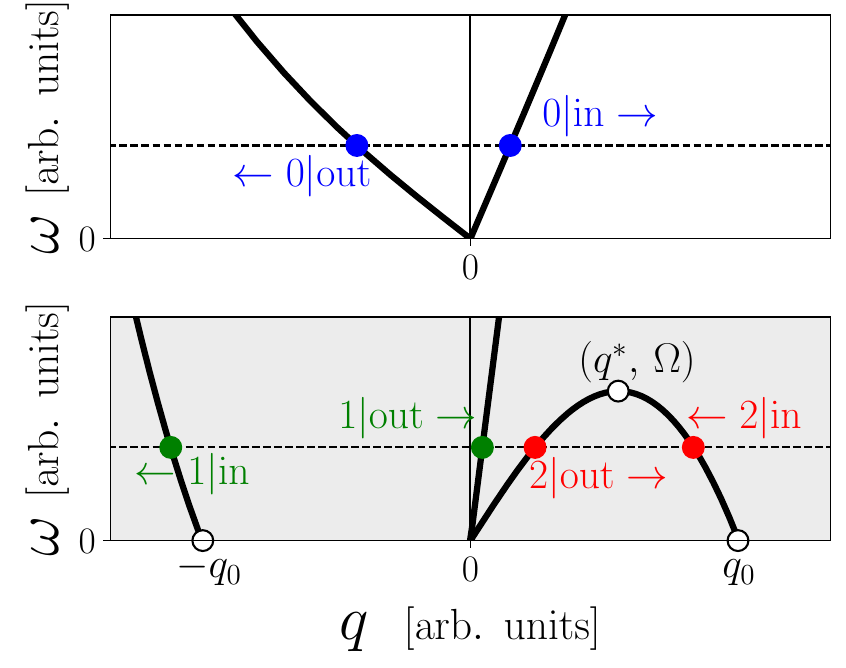}
\caption{Graphical representation of the positive frequency part of
  the dispersion relation \eqref{eq.excit1} in the far upstream
  subsonic (upper plot) and downstream supersonic (lower plot)
  regions. The background color of the lower plot is greyed for
  recalling that it concerns the interior of the analogue black
  hole. In both plots the horizontal dashed line represents the
  angular frequency $\omega$ of a given excitation. In the upstream
  region there are two channels of propagation associated to each
  value of $\omega$. In the downstream region there are four (two)
  propagation channels when $\omega$ is smaller (larger) than the
  threshold $\Omega$ defined in Eq. \eqref{threshold}. The channels
  are denoted as 0, 1 or 2, with an additional ``in'' (``out'') label
  indicating if the wave propagates towards (away from) the
  horizon. The direction of propagation of each channel is marked with
  an arrow.}\label{fig2}
\end{figure}

The decomposition \eqref{eq.model2} is meaningful in a regime of small quantum fluctuations where the operator $\hat{\psi}$ can be treated within a Bogoliubov approach. In this case  $\hat{\psi}$ is naturally expanded along the asymptotic ingoing and outgoing channels of the flow. The dispersion relation of elementary excitations in the asymptotic upstream subsonic and downstream supersonic regions 
($x\to -\infty$ and $+\infty$, respectively) 
is of Bogoliubov type, with a Doppler shift accounting for the finite velocity of the background:
\begin{equation}\label{eq.excit1}
\left(\omega - V_\alpha q\right)^2=\omega_{{\sss\rm B},\alpha}^2(q),
\end{equation}
where $\alpha=u$ far upstream and $\alpha=d$ downstream. In this expression  $V_\alpha$ is the asymptotic velocity of the flow in region $\alpha$ and
\begin{equation}\label{eq.excit2}
\omega_{{\sss\rm B},\alpha}(q)=c_\alpha q \sqrt{1+q^2\xi^2_\alpha/4}
\end{equation}
is the Bogoliubov dispersion relation \cite{Bogoliubov1947}. In Eq. \eqref{eq.excit2} $c_\alpha$  and $\xi_\alpha$ are the asymptotic speed of sound and healing length in region $\alpha$, respectively, see Appendix \ref{app.different}.
These dispersion relations are represented in Fig. \ref{fig2}.

From the identification of the relevant channels and of their
direction of propagation it is possible to define quantum modes
forming a basis enabling to describe all the elementary excitations of
the background flow. To each such mode is associated a quantum
operator: $\hat{b}_i^\dagger(\omega)$ and $\hat{b}_i(\omega)$ ($i=0$,
1 or 2) are the creation and annihilation operators of an excitation
of energy $\hbar\omega$ which is ingoing in channel $i|{\rm in}$ and
scattered by the horizon onto the three outgoing channels
$0|{\rm out}$, $1|{\rm out}$ and $2|{\rm out}$. Since each $\hat{b}$
mode is associated with a single ingoing channel, it is denoted as an
``ingoing mode''. It is also relevant to define ``outgoing modes''
associated with a single outgoing channel. The corresponding operators
are denoted as $\hat{c}_i(\omega)$ and
$\hat{c}_i^\dagger(\omega)$. For instance $\hat{c}^ \dagger_0$ is the
creation operator of an excitation where the three ingoing channels
are implied and form an outgoing excitation in channel $0|{\rm out}$.
The corresponding quantum mode is the analogous Hawking mode. The
outgoing modes are related to the incoming ones via the scattering
matrix $S(\omega)$:
\begin{equation} \label{eq.excit3}
\begin{pmatrix}
\hat{c}_{0}  \\
\hat{c}_{1} \\
\hat{c}_{2}^{\dagger} \\
\end{pmatrix}
=
\begin{pmatrix}
S_{00} & S_{01} & S_{02} \\
S_{10} & S_{11} & S_{12} \\
S_{20} & S_{21} & S_{22} \\
\end{pmatrix}
\,
\begin{pmatrix}
\hat{b}_{0}  \\
\hat{b}_{1} \\
\hat{b}_{2}^{\dagger} \\
\end{pmatrix},
\end{equation}
where all the $\omega$ dependencies have been omitted for legibility. 
The modes $\hat{b}_2$ and $\hat{c}_2$ are particular in the sense that they have a negative norm and should be quantized inverting the usual role of the creation and annihilation operators \cite{Blaizot1986} in order to satisfy the standard Bose commutation relations. The mode $\hat{c}_2$ is analogous to what is called the partner is the context of Hawking radiation. We denote the mode associated with $\hat{c}_1$ the companion; Lorentz invariance prevents such a mode to exist in a gravitational black hole, but it is unavoidable in an analog system.

The fact that the outgoing operators fulfil the canonical commutation relations implies that the scattering matrix $S(\omega)$ obeys the skew-unitarity condition
\begin{equation}\label{eq.excit4}
    S^\dagger \eta S =\eta =S \eta S^\dagger ,
    \quad\mbox{where}\quad
    \eta={\rm diag}(1,1,-1)\; .
\end{equation}
For $\omega>\Omega$ the mode with subscript 2 (the partner) disappears because the corresponding ingoing and outgoing channels do (cf.~Fig.~\ref{fig2}) and the $S$-matrix becomes $2\times 2$ and unitary. In this case the vacuum of the outgoing modes (the $\hat{c}$'s) is identical to the vacuum of the incoming ones (the $\hat{b}$'s) and the analogous Hawking effect disappears.
The value of the corresponding threshold energy is
\begin{equation}\label{threshold}
\begin{split}
& \Omega=q^* V_d-\omega_{{\sss\rm B},d}(q^*) \quad\mbox{with}\\
& q^*\xi_{d}=\left(-2+\frac{m_d^2}{2}+
\frac{m_d}{2}
\sqrt{8+m_d^2}\right)^{\frac{1}{2}}.
\end{split}
\end{equation}

\begin{figure}
\centering
\includegraphics[width=\linewidth]{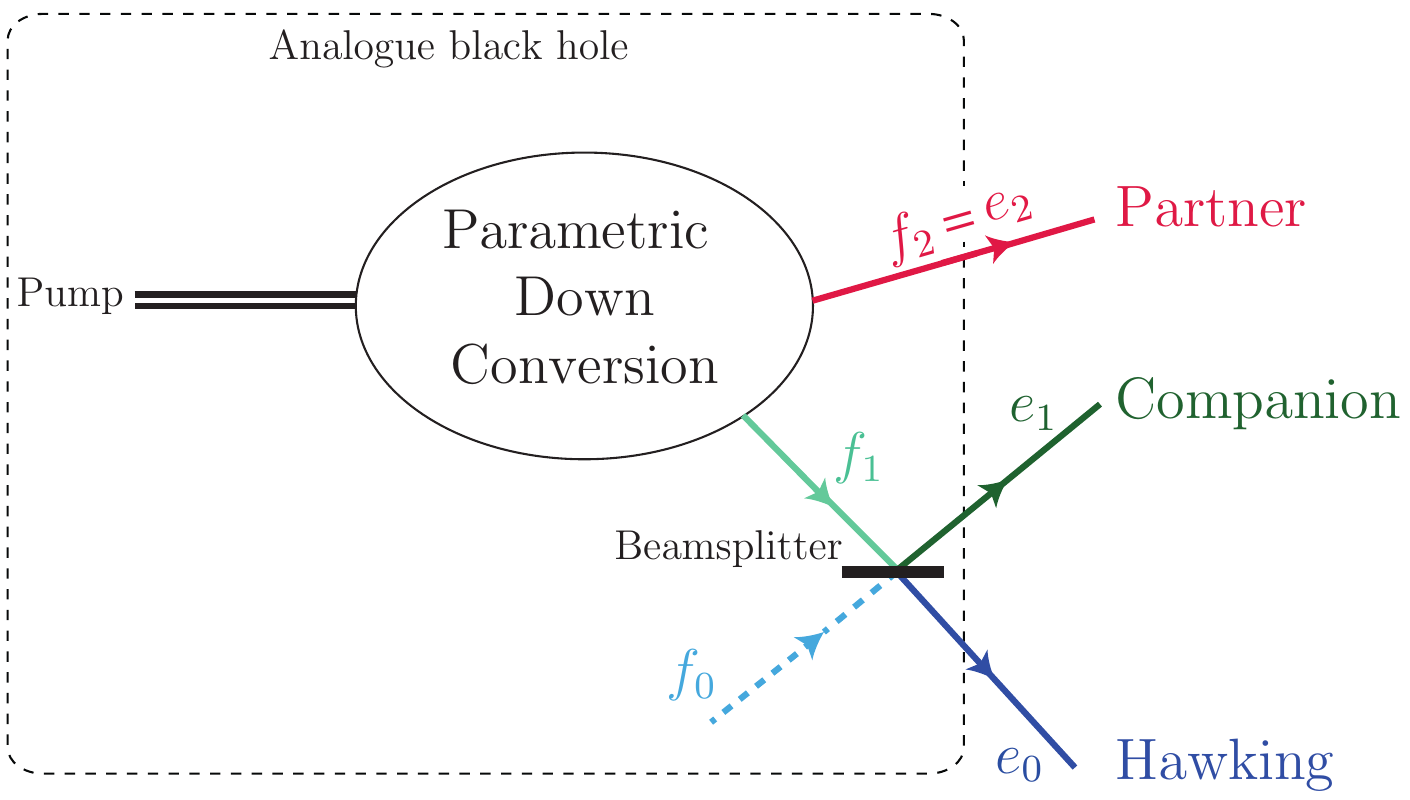}
\caption{Model optical system equivalent to the analogous black
  hole. The explicit relationship of the effective modes $f_i$ and
  $e_i$ ($i=0,1$ and 2) with the physical outgoing modes is given in
  Eqs. \eqref{eq.apsig9} and \eqref{eq.apsig10}. The $f_0$ mode is
  represented with a dashed line because, contrarily to modes $f_1$
  and $f_2$, it is not occupied at zero temperature, see
  Eqs. \eqref{eq.apsig12}. The long wavelength transmission
  coefficient of the beam splitter is denoted as $\Gamma_0$ in
  Appendix \ref{app.sigma}. It plays the role of the grey body factor
  of the analogous black hole.}\label{fig.optical.model}
\end{figure}

It has been shown in Ref. \cite{Isoard2021} that the three-mode system describing the analogous black hole horizon can be modeled by an optical setup simply composed of a parametric amplifier and a beam splitter, as depicted in Fig. \ref{fig.optical.model}.
Entanglement is localized in the two-mode squeezed state $f_1|f_2$. It is then dispatched by mean of a beam splitter which performs a non local transformation from the effective modes $(f_0,f_1,f_2)$
onto other effective modes $(e_0,e_1,e_2)$ which connect to the physical outgoing modes $(c_0,c_1,c_2)$ by a simple local linear unitary Bogoliubov 
transformation (LLUBO), see details in Appendix \ref{app:analogue}. We stress that the configuration depicted in Fig. \ref{fig.optical.model} captures the essence of the analogue black hole configuration and is generic: A similar model has been used in Ref. \cite{Agullo2022} for describing an optical system containing a pair white-black hole analog and then studied on general grounds in \cite{Brady2022}.
Albeit the fact that we allow the down-conversion process to violate Lorentz invariance (since we 
enforce the Bogoliubov dispersion relation) the two-mode squeezed state built on $\hat{f}_1$ and $\hat{f}_2$ is the closest possible analogue to an ideal black hole. The additional beam splitter is inherent to any analogue model. The transmission coefficient of this beam splitter plays the role of the grey-body factor which is relevant in the gravitational context, see discussion in \cite{Isoard2021}.

\subsection{Density matrix and measure of bipartite entanglement}\label{subsec.DM}

For describing analog black holes physics one should consider observables operating in the Fock space of the outgoing modes, as would 
an observer located outside the horizon of a gravitational black hole. However, the physical implementation of the analog black hole configurations
presented in Appendix \ref{app.different}
is realized in the vacuum  of the incoming modes.
This mismatch is the origin of the quantum evaporation process, as first presented by Hawking \cite{Hawking1975}.
An ingoing vacuum mode of frequency $\omega$ (which we denote as $|0_\omega\rangle^{\rm in}$) relates to an outgoing vacuum mode $|0_\omega\rangle^{\rm out}$ 
through
\begin{equation} \label{eq.entang1}
|0_\omega\rangle^{\rm in} = \frac{1}{|S_{22}|} \,  
e^{\left(X_{02} \, \hat{c}_{0}^\dagger  + X_{12} \, \hat{c}_{1}^\dagger \right) \, \hat{c}_{2}^\dagger} \,  |0_\omega\rangle^{\rm out},
\end{equation}
where the coefficients $X_{i2}$ ($i=0$ and 1) are expressed in terms of entries of the $S$-matrix \eqref{eq.excit3}:
$X_{i2}(\omega)=S_{i2}(\omega)/S_{22}(\omega)$,
and
\begin{equation}\label{eq.3modes0omega}
|0_\omega\rangle^{\rm out}=|0_\omega\rangle_0^{\rm out}\otimes
|0_\omega\rangle_1^{\rm out} \otimes |0_\omega\rangle_2^{\rm out},
\end{equation}
where $|0_\omega\rangle_j^{\rm out}$ is the vacuum
of operator $\hat{c}_j(\omega)$, with $j\in\{0,1,2\}$. 
 Eq. \eqref{eq.entang1} can be rewritten in terms of the number states of quasi-particles of type $c_j$ 
\begin{equation} \label{eq.entang2}
 |n_\omega\rangle_j=\frac{1}{\sqrt{n!}}\left(\hat{c}_j^\dagger(\omega)\right)^n
 |0_\omega\rangle_j^{\rm out},
     \end{equation}
as
\begin{equation} 
\label{eq.entang3}
|0\rangle^{\rm in} = \frac{1}{|S_{22}|} \sum_{\mu,\nu=0}^{\infty} 
\sqrt{\binom{\mu+ \nu}{\mu}} X_{\sss 02}^\mu X_{\sss 12}^{\nu}
|\mu\rangle_{\sss 0}|\nu\rangle_{\sss 1}|\mu+\nu\rangle_{\sss 2}.
\end{equation}
It will be often the case below that, as in expression \eqref{eq.entang3}, we drop 
the explicit $\omega$-dependence for legibility. 

Relationships \eqref{eq.entang1} and  \eqref{eq.entang3} ensure that the system is in a pure three modes Gaussian state which is fully characterized by its $6\times 6$ covariance matrix $\sigma(\omega)$ with entries
\begin{equation}\label{eq.entang4}
  \sigma_{\ell m} \equiv \frac{1}{2} \, \langle \hat{\xi}_\ell \, \hat{\xi}_m +
  \hat{\xi}_m \, \hat{\xi}_\ell  \rangle - \langle \hat{\xi}_\ell \rangle
  \, \langle \hat{\xi}_m \rangle,
\end{equation}
where operator $\hat{\xi}_\ell$ (or $\hat{\xi}_m$) is one of the 6 components of the column vector 
$\hat{\boldsymbol{\xi}} = \sqrt2 \,
(\hat{q}_0,\hat{p}_0,\hat{q}_{1},\hat{p}_{1},\hat{q}_{2},\hat{p}_2)^{\rm \sss T }$, where 
\begin{equation}\label{eq:qetp}
 \hat{q}_j(\omega)=\frac{1}{\sqrt{2}}(\hat{c}_j+\hat{c}_j^\dagger)\quad \mbox{and}\quad
 \hat{p}_j(\omega)=\frac{i}{\sqrt{2}}(\hat{c}_j^\dagger-\hat{c}_j).
\end{equation}
In expression \eqref{eq.entang4} and in all the following, the
averages $\langle \cdots \rangle$ are performed over the density
matrix of the system.  This density matrix is simply
$\rho=|0 \rangle^{\rm in}\, ^{\rm in}\langle 0|$ in the ideal case just described.
We also consider more realistic situations where some incoherent
excitations are present and the system is not in a pure state. A
simple manner to account for this situation would be to assume that
the system is in a thermal state. This is however impossible because
the analog configurations depicted in Appendix \ref{app.different}
are thermodynamically unstable. A way to circumvent this problem has
been proposed in Refs. \cite{Macher2009,Recati2009}. It consists in postulating
that the system was initially in thermal equilibrium at temperature $T$
with a constant density and velocity ($n_u$ and $V_u$, respectively)
and that the flow has been adiabatically modified by slowly ramping the
appropriate external potential, eventually reaching the black 
hole configuration
of interest. This situation, although idealized, it less schematic
than the zero excitation regime. It emulates the experimental
situation of Refs. \cite{De_nova_2019,Kolobov_2021} if the system is
considered to have been initially in equilibrium in the frame attached to the
flowing condensate. In this case one has (for $j\in\{0,1,2\}$)
\begin{equation}\label{eq.entang5}
\bar{n}_j(\omega)\equiv 
\langle \hat{b}^\dagger_j(\omega) \hat{b}_j(\omega) \rangle
= n_{\rm \sss th}\!
\left[ \omega_{{\sss \rm B},\alpha} \!\left(  q_{j|\rm in}(\omega)  \right) \right],
\end{equation}
where $n_{\rm \sss th}(\omega)$ is the thermal Bose occupation
distribution at temperature $T$ and energy $\hbar\omega$, whereas
$\omega_{{\sss \rm B},\alpha}(q_{j|\rm in})$ is the Bogoliubov
dispersion relation \eqref{eq.excit2}, with $\alpha = u$ if $j=0$ and
$\alpha=d$ if $j=1$ or 2. The functions $q_{j|\rm in}(\omega)$
appearing in expression \eqref{eq.entang5} are pictorially defined in
Fig. \ref{fig2}. For instance $q_{2|\rm in}(\omega)$ is the function
that, to a given angular frequency $\omega\in[0,\Omega]$, associates a
wave-vector along the $2|\rm in$ dispersion branch\footnote{
  $q_{2|\rm in}(\omega)\in[q^*,q_0]$ with $q^*=q_{2|\rm in}(\Omega)$
  and $q_0=q_{2|\rm in}(0)$, see Fig. \ref{fig2}.}. We loosely refer
to the cases where $\bar{n}_j$ ($j\in\{0,1,2\}$) is equal to the
right-hand side (r.h.s.) of \eqref{eq.entang5} as ``finite
temperature'' situations. In the simplest configuration, denoted as
``zero temperature'' the system is in the pure state
$|0\rangle^{\rm in}$ and the $\bar{n}_j$'s are all equal to zero.  We
will not specify the values of the $\bar{n}_j$'s in the following, so
that the formulas we give are generally valid, even in situations
where the occupation numbers should not be given by formulas of the
type of Eq. \eqref{eq.entang5}. However, for illustrative purposes,
all the figures of the paper are plotted in specific temperature
cases.

Several observables have been proposed to theoretically evaluate the
bipartite entanglement in the context of analogue gravity, such as the
Cauchy-Schwarz criterion
\cite{deNova2014,Busch2014a,Boiron2015,deNova2015,Steinhauer2015,Coutant2018},
the generalized Peres-Horodecki parameter
\cite{Finazzi2014,Robertson2017}, the logarithmic negativity
\cite{Horstmann_2011,Jacquet2020,Agullo2022b,Delhom2023}, the
entanglement entropy \cite{Giovanazzi2011,Delhom2023}, the
entanglement of formation \cite{Bruschi_2013}, the Gaussian contangle
\cite{Isoard2021}.  In the present work, for reasons recalled in
Appendix \ref{app.sigma}, we chose  as in Ref. \cite{Isoard2021} to evaluate the bipartite
entanglement between modes $i$ and $j$ by means of a quantity
$\Lambda^{(i|j)}(\omega) \in ]-\infty,1]$ which is a monotonous
measure of entanglement that we denote as the ``PPT measure''.  States $i$ and $j$ are separable when
$\Lambda^{(i|j)}<0$, which is always the case when $(i|j)=(0|1)$: the
companion and the Hawking modes are not entangled. The two other
couples of modes, $(0|2)$ and $(1|2)$, are always entangled at $T=0$
for all $\omega$. Their entanglement decreases with increasing
temperature by an amount specified by the PPT measure which reads
explicitly ($i=0$ or 1) \cite{Isoard2021}:
\begin{equation}
\begin{split}\label{eq.entang6}
    \Lambda^{(i|2)}(\omega)= & 
    - \langle \hat{c}_i^\dagger \, \hat{c}_i \rangle
    - \langle \hat{c}_2^\dagger \, \hat{c}_2 \rangle
    \\
    & +
    \sqrt{
    \left(\langle \hat{c}_i^\dagger \, \hat{c}_i \rangle
    - \langle \hat{c}_2^\dagger \, \hat{c}_2 \rangle\right)^2 +
    4 |\langle \hat{c}_i \, \hat{c}_2 \rangle|^2
    }\; .
\end{split}
\end{equation}
As clear from the above expression, a key ingredient for characterizing entanglement is the determination of average values of different combination of two creation or annihilation operators of the outgoing 
modes. At the experimental level, this determination could reveal difficult for quantities such as $\langle \hat{c}_0 \, \hat{c}_2 \rangle$ for instance. Steinhauer proposed a possible way to extract this information from the knowledge of the density-density correlation function \cite{Steinhauer2015}. As stressed in Refs. 
\cite{Isoard2020,Isoard_PhD} this method needs to be used with more care than initially thought, but is indeed a possible 
manner to obtain the information. At the theoretical level, it is a straightforward matter to compute the expectation values of products of the ingoing creation and annihilation operator (quantities such as \eqref{eq.entang5} for instance). From there, expression \eqref{eq.excit3} makes it possible to compute the equivalent expressions for the 
outgoing operators, which are the quantities of interest. The relevant formulae are given in Appendix \ref{app.sigma}, Eqs. \eqref{eq.apsig8}.

\section{Bipartite nonlocality}\label{sec.Bell2}

Most Bell-like inequalities proposed in the context of continuous variables hinge on a discretization process \cite{Leonhardt1995,Gilchrist1998,Auberson2002, Wenger2003,Garc2004,Martin2017}. Indeed, since the set of outcomes for a given observable is typically unbounded in continuous variable systems, it seems \textit{a priori} difficult to derive an upper bound of the expectation value of a Bell-type observable. It is not impossible, though: relying on the  Fine–Abramsky–Brandenburger theorem \cite{Fine1982,Abramsky2011}, 
the authors of Refs. \cite{Cavalcanti2007,Barbosa2022} were able to derive continuous Bell inequalities for continuous and unbounded observables.
However, to our knowledge, no practical (theoretical or experimental) use of such fully continuous approaches has been successfully implement so far.
A method to circumvent this difficulty is based on discretization and consists of using a dichotomic binning of the outcome results. In this case, by using observables which rely on continuous measurements but can only take a finite number of outcomes, one can construct a Bell inequality similar to those derived for discrete variables. A well-known example of such observables are the so-called pseudo-spin operators: they live in an infinite-dimensional Hilbert space  but the outcome of their measurement is either $-1$ or $+1$.

Due to its practicability, we chose to follow this discretization approach to derive Bell-like inequalities in the context of BEC
analogue black holes. In this work we use the GKMR (Gour, Khanna, Mann and Revzen) pseudo-spins introduced in Ref. \cite{Gour2004}:
To an outgoing mode $j$ ($j\in\{0,1,2\}$) of energy $\omega$ is associated a Hermitian vectorial operator $\hat{\boldsymbol{\Pi}}{}^{(j)}(\omega)$ with Cartesian coordinates
\begin{subequations}\label{eq.GKMR1}
\begin{align}
& \hat{\Pi}^{(j)}_x(\omega) = \int_0^{+\infty} {\rm d}q \, \Big(
|q\rangle_j\,_j\langle q|-|-q\rangle_j\,_j\langle -q|
\Big), \label{eq.GKMR1a}  \\
& \hat{\Pi}^{(j)}_y(\omega) = i \int_0^{+\infty} {\rm d}q\, \Big(
|q\rangle_j\,_j\langle -q|-|-q\rangle_j\,_j\langle q|
\Big), \label{eq.GKMR1b} \\
& \hat{\Pi}^{(j)}_z(\omega) = \int_{-\infty}^{+\infty} {\rm d}q \,
|q\rangle_j\,_j\langle -q|,\label{eq.GKMR1c}
\end{align}
\end{subequations}
where $|q\rangle_j$ is the eigenstate  associated to the eigenvalue $q$ of the position operator $\hat{q}_j(\omega)$ \eqref{eq:qetp}.
The operators \eqref{eq.GKMR1} anti-commute with each others and all square to unity. They verify the expected spin commutation relations, such as
\begin{equation}\label{eq.GKMR2}
    \left[\hat{\Pi}_x^{(j)},\hat{\Pi}_y^{(j)}\right]=2 \, i\, \hat{\Pi}_z^{(j)},
\end{equation}
and similar relations upon circular permutations of the indices $x$, $y$ and $z$. 
In Appendix \ref{app.pseudo.spin} we recall the properties of the pseudo-spin operator and of its eigenvalues which are useful in the following.  

These pseudo-spin operators have been studied in contexts similar to ours in Refs. \cite{Martin2017,Xiang2017,Chatterjee2022}.
Compared with other pseudo-spin operators such as for instance those introduced by Banaszek and Wodkiewicz
\cite{Banaszek1999,Chen2002}, the GKMR spins \eqref{eq.GKMR1} have the advantage of having simple Wigner transforms, which makes the computation of their expectation values over Gaussian states relatively easy, as detailed in Appendix \ref{app.averages}. 
We stress here that different choices of spin representation lead to different values of averages of the Bell operators \eqref{eq.GKMR3} \cite{Gour2004}. From this remark naturally results that, generally speaking, the observables \eqref{eq.GKMR6}, \eqref{eq.TP7} and \eqref{eq.M1}  we use below are witnesses of nonlocality: Their violation of Bell-type inequalities is a sufficient but not necessary test of nonlocal behavior.
As a final remark, we note here an unforeseen benefit of the use of the pseudo-spin operators \eqref{eq.GKMR1}: it will be shown in  Sec. \ref{sec.Bell3} that the structure 
of the zero temperature ground state of the analog system (the vacuum $|
0\rangle^{\rm in}$)
is most easily analyzed in terms of a combination of eigenstates of operator $\hat{\Pi}_x$.

Equipped with the pseudo-spin operators \eqref{eq.GKMR1} we can define a CHSH Bell operator \cite{Clauser1969} measuring the correlations between the emitted quasi-particles of type $i$ and $j$:
\begin{equation}\label{eq.GKMR3}
\begin{split}
  \hat{\mathscr{B}}{}^{(i|j)}(\omega) = & (\boldsymbol{a}+ \boldsymbol{a}')\cdot\hat{\boldsymbol{\Pi}}{}^{(i)}
   \otimes \boldsymbol{b}\cdot\hat{\boldsymbol{\Pi}}{}^{(j)} \\
   + &
(\boldsymbol{a}-\boldsymbol{a}')\cdot\hat{\boldsymbol{\Pi}}{}^{(i)}
   \otimes \boldsymbol{b}'\cdot\hat{\boldsymbol{\Pi}}{}^{(j)} , 
\end{split}
\end{equation}
where $\boldsymbol{a}$, $\boldsymbol{a}'$, $\boldsymbol{b}$ and $\boldsymbol{b}'$ are unit vectors.
Given a unit vector $\boldsymbol{n}$ it is easily checked that $(\boldsymbol{n}\cdot\hat{\boldsymbol{\Pi}}{}^{(j)})^2=\mathbb{1}$, meaning that the Hermitian operator $\boldsymbol{n}\cdot\hat{\boldsymbol{\Pi}}{}^{(j)}$ has eigenvalues $\pm 1$.
It then follows from direct inspection that, if one believes in local realism, one should expect that a measure of the operator $\hat{\mathscr{B}}{}^{(i|j)}$ yields a result $\pm 2$ \cite{rem_local}.
The tenant of local realism is thus violated when $\langle \hat{\mathscr{B}} {}^{(i|j)}\rangle>2$, whereas Cirel’son bound \cite{Cirelson1980} imposes
$\langle \hat{\mathscr{B}}{}^{(i|j)} \rangle\le 2\sqrt{2}$.  Since the modes 0 and 1 are not entangled the quantity $\langle \hat{\mathscr{B}}{}^{(0|1)}\rangle$ is always lower than 2 and its computation is of no interest to us. For attempting to violate as much as possible Bell inequality one should consider the modes 0 and 2 (or 1 and 2) and look for an arrangement of the four measurement directions $\boldsymbol{a}$, $\boldsymbol{a}'$, 
$\boldsymbol{b}$ and $\boldsymbol{b}'$ which maximizes $\langle \hat{\mathscr{B}}{}^{(i|2)} \rangle$ ($i=0$ or 1). This procedure is explained in Appendix \ref{app.maxi2} and makes it possible to analytically compute the quantity 
\begin{equation}\label{eq.GKMR6}
B^{(i|2)}(\omega)\equiv \, \mbox{max}_{\boldsymbol{a},\boldsymbol{a}',\boldsymbol{b}, \boldsymbol{b}'}
\left\langle \hat{\mathscr{B}}{}^{(i|2)} (\omega) \right\rangle.
\end{equation}
The corresponding explicit expression is given in Eq. \eqref{GoodCHSHb}. 

\begin{figure}[h]
\centering
\includegraphics[width=\linewidth]{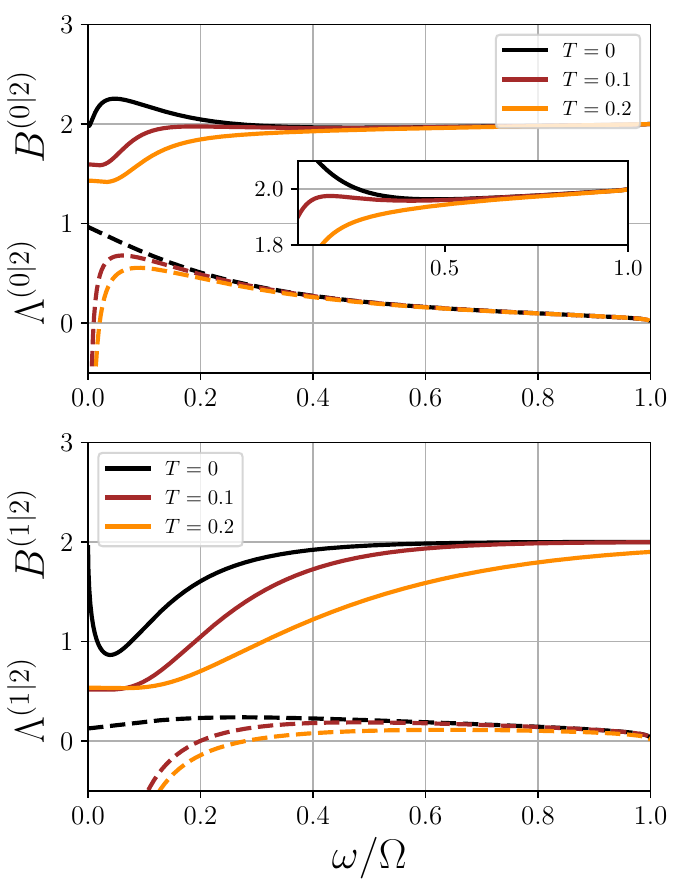}
\caption{Plot of $B^{(i|2)}$ (solid lines) and $\Lambda^{(i|2)}$
  (dashed lines) as functions of $\omega$ for the waterfall
  configuration with $m_d=2.9$ and different temperatures. We only
  consider the range of frequencies $\omega<\Omega$ for which the
  vacua of the outgoing and ingoing modes do not coincide. The value
  of the temperature is indicated in units of $g\, n_u=m c_u^2$. Upper
  plot: $i=0$, lower plot $i=1$. Non separability of modes $i$ and $2$
  is achieved when $\Lambda^{(i|2)}>0$. Bell inequality is violated
  when $B^{(i|2)}>2$. The inset in the upper plot is a blow-up of the region $1.8\le B^{(0|2)}\le 2.1$ and $0.1\le \omega/\Omega\le 1$ showing that the reduced state $(0|2)$ does not violate Bell inequality at temperatures $T=0.1$ and 0.2.}
\label{fig3.1}
\end{figure}

The values of $B^{(0|2)}$ and $B^{(1|2)}$ are plotted as functions of $\omega$ for different temperatures in Fig. \ref{fig3.1} for a waterfall configuration with a downstream Mach number\footnote{In this theoretical model configuration, fixing the value of $m_d$ determines all the other dimensionless parameters: $m_d=m_u^{-2}=n_u/n_d=V_d/V_u$, see Ref. \cite{Larre2012}.} 
$m_d\equiv V_d/c_d=2.9$, the same as in the Technion 2019 experiment \cite{De_nova_2019}. We also plot for comparison the values of the 
corresponding PPT measures  $\Lambda^{(0|2)}$  and $\Lambda^{(1|2)}
$, as defined by \eqref{eq.entang6}. The figure illustrates that, as well known, 
entanglement is necessary but not sufficient for violating Bell inequality. Also, the amount by which the Bell inequality is 
violated is not proportional to the amount of entanglement. This is clearly seen, for instance, by comparing the values of 
$B^{(0|2)}$ and $\Lambda^{(0|2)}$ at $T=0$: the maximum violation of Bell inequality is not achieved for the maximal entanglement. 
One can also notice that the violation of Bell inequality is much less resilient to temperature than is the entanglement. These 
features can be most easily understood in the framework of the optical model represented in Fig. \ref{fig.optical.model}. They originate from the
the dilution of entanglement between the three modes caused by the beam splitter, as discussed in Appendix \ref{app:analogue}.

\begin{figure}
    \centering
    \includegraphics[width=\linewidth]{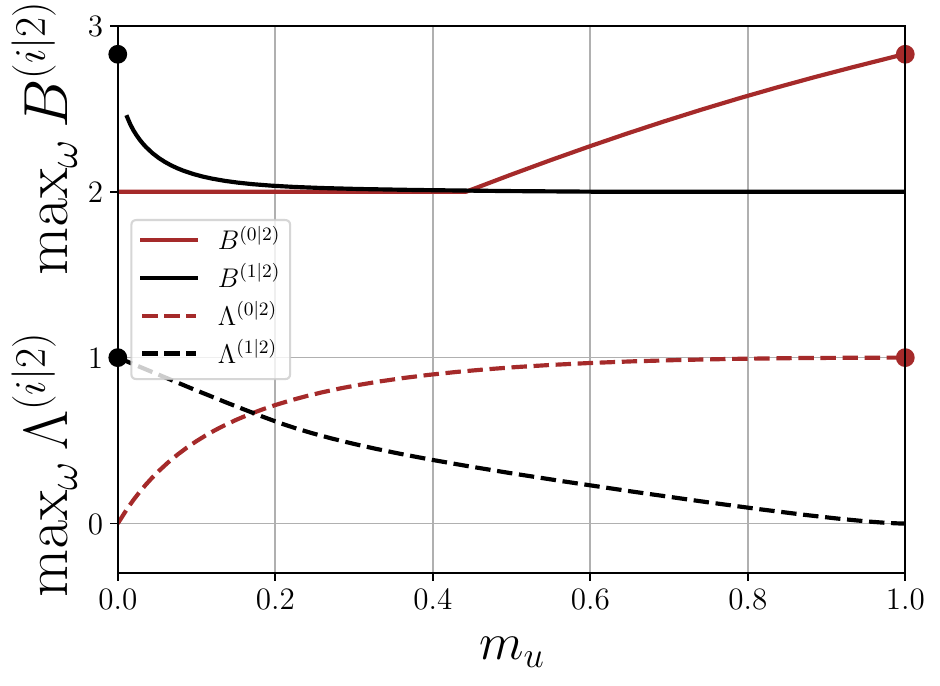}
    \caption{Zero temperature value of the CHSH parameter and of the
      PPT measure characterizing non separability of modes $i$ ($=0$
      and 1) and 2 for the waterfall configuration. The maximal value
      reached by these quantities over the interval
      $\omega\in[0,\Omega]$ is plotted as a function of the upper Mach
      number $m_u$ which (as explained in Appendix
      \ref{app.different}) characterizes a given configuration. The
      values of $B^{(1|2)}$ for $m_u\le 0.01$ are not indicated
      because of lack of numerical precision. The upper bounds of
      $B^{(i|2)}$ and $\Lambda^{(i|2)}$ ($\sqrt{8}$ and 1,
      respectively) are indicated with filled dots.}
    \label{fig3:plus}
\end{figure}

Fig. \ref{fig3.1} indicates that, in the waterfall configuration we consider (with $m_d=2.9$),
the entanglement is lower and the violation of Bell inequality less significant for the correlations among 
modes 1 and 2 than for the correlations among modes 0 and 2. However, this is not always the case. This point is illustrated in Fig. \ref{fig3:plus}  which displays, for all waterfall configurations, the zero 
temperature values of ${\rm max}_\omega B^{(i|2)}$ and ${\rm max}_\omega \Lambda^{(i|2)}$ (for 
$i=0$ and 1) as functions of the upper Mach number $m_u$ (the maximisation 
is performed at fixed $m_u$, for $\omega\in[0,\Omega]$). All the possible waterfall 
configurations are considered since $m_u$ spans the whole interval $[0,1]$. In this figure we aim 
at evaluating the largest amount of entanglement and nonlocality reached in each configuration. This is the reason why
we plot the maximum values taken by the quantities 
$\Lambda^{(i|2)}$ and $B^{(i|2)}$ over the energy interval $[0,\Omega]$ [where $\Omega$ is defined 
by Relation \eqref{threshold}] since this is only for energies in this interval that spontaneous emission of 
quasi-particles occurs.
For instance, the situation depicted in Fig. \ref{fig3.1} ($m_d=2.9$) corresponds in Fig. \ref{fig3:plus} to the point $m_u=0.587$ since for waterfall configurations $m_u$ and $m_d$ are related by 
\eqref{eq.model4a}. And indeed, Fig. \ref{fig3:plus} shows that for this value of $m_u$ the maximum over $\omega$ of $B^{(0|2)}$ is 2.25, and the one of $B^{(1|2)}$ is 2, as observed in Fig. \ref{fig3.1}.

\begin{figure}
    \centering
    \includegraphics[width=\linewidth]{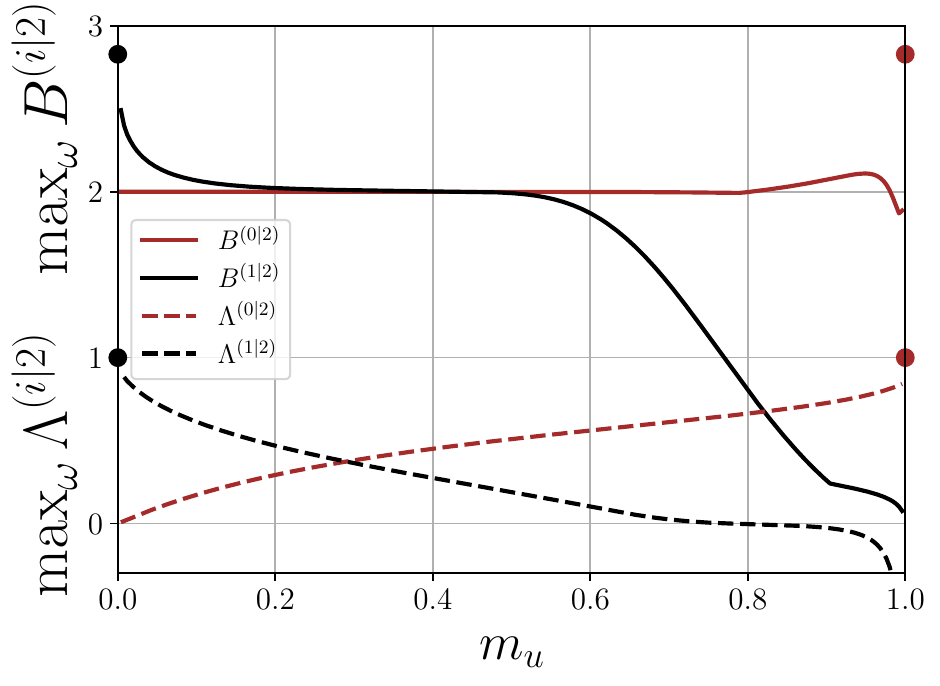}
    \caption{Same as Fig. \ref{fig3:plus} for a temperature
      $T=0.2\, g n_u$. Contrarily to what is observed in the zero
      temperature case displayed in Fig. \ref{fig3:plus}, the maxima
      of $\Lambda$ and $B$ (1 and $2\sqrt{2}$, respectively) are never
      reached.}
    \label{fig3.pluT}
\end{figure}

Fig. \ref{fig3:plus} shows that for values of $m_u$ larger than 0.6, the entanglement is mainly concentrated between modes 2 and 0, 
i.e., between the Hawking quantum and the partner. This is indicated by the fact that both the PPT measure and the CHSH parameter significantly point to nonseparability and nonlocality between these two modes. 
For $m_u\le 0.2$ instead, the figure shows that entanglement is concentrated between the partner and the companion (modes 2 and 1, respectively). 
A plot similar to the one of Fig. \ref{fig3:plus}, but where the
quantities are evaluated at finite temperature, enables to evaluate
the resilience of entanglement and non-separability to an increase of
temperature.  This check is performed in Fig. \ref{fig3.pluT} which
shows that, whereas at $T=0.2\, gn_u$ the PPT measure is not
dramatically affected, the CHSH parameters $B^{(0|2)}$ and $B^{(1|2)}$
no longer show evidences of violation of Bell inequality, except in
the $(1|2)$ sector for waterfall configurations with
$m_u\lesssim 0.15$ and, in a lesser extent, in the $(0|2)$ sector for
$m_u\gtrsim 0.85$.

Figs. \ref{fig3:plus} and \ref{fig3.pluT} illustrate a specific feature of analog models: there is no recipe enabling to qualify one of the positive norm modes (in our specific case, mode 0 or 1) as 
unessential. For instance, it is incorrect to study the system discarding {\it a priori} the companion (mode 1) from one's analysis. Fig. \ref{fig3:plus} shows that, at zero temperature, this is allowed in some regions of parameters, but incorrect in others. Fig. 
\ref{fig3.pluT} even shows that only the companion-partner correlations display 
violation of Bell inequality above a certain temperature. In this instance it is the Hawking mode which is unessential. 
It is therefore important to give a proper account of all the modes involved in our analog system.

Another way to consider the same problem is to use the equivalent model depicted in Fig. \ref{fig.optical.model}: it is well known that a 
nondegenerate optical parametric amplifier generates an EPR state when the squeezing parameter tends to infinity (see, e.g., Ref. \cite{Banaszek1999a}). As discussed in Appendix \ref{app:analogue} this is the case for the two mode squeezed state \eqref{eq.apsig11bis} when $\omega\to 0$.
The zero energy transmission coefficient $\Gamma_0$ of the effective beam splitter 
(pictorially defined in  Fig. \ref{fig.optical.model}) tends to 0 or 1 when the upper Mach number $m_u$ tends to 0 or 1, respectively 
[see e.g., Eq. \eqref{eq.Gamma0} which holds for the waterfall 
configuration]. In these 
two limits the system forms an EPR state between two of the three outgoing modes (and the third one can be omitted). This EPR state involves, either the Hawking quantum 
and the partner (when $m_u \to 1$) or the companion and the partner (when $m_u \to 0)$.
This is the reason why the entanglement and nonlocality bounds are reached in this two limits in Fig. \ref{fig3:plus}, and similarly in Figs. \ref{fig.H1} and \ref{fig.H2} of Appendix \ref{app.other} for the delta peak and flat profile configurations, respectively.

We conclude this section by noticing that, from a quantum information perspective,  
the fact that reduced bipartite states of a tripartite system are entangled 
is of no particular significance {\it per se}. It is however important for future experimental studies of analog systems to determine for which configurations, and to quantify to which extent, the three-mode
acoustic Hawking emission is bipartite
entangled and nonlocal. Besides, we will see in the next section that this resilience of entanglement to partial tracing acquires a particular significance when examining the exact nature of the long wavelength components of the three-mode state $|0\rangle^{\rm in}$ (the ground state of the system).

\section{tripartite nonlocality} \label{sec.Bell3}

Equipped with the same pseudo-spin operators as the ones defined in Sec. \ref{sec.Bell2} one can define a three-mode Bell operator of a type similar to the two-mode one \eqref{eq.GKMR3}. This operator measures the correlations between the outgoing quasi-particles of type $i$, $j$ and $k$. It is defined as  \cite{Svetlichny1987,Bancal2011}:
\begin{equation}\label{eq.TP0}
\hat{\mathscr{S}}{}^{(i|j|k)}(\omega)  =  
 \tfrac12 \hat{\mathscr{B}}{}^{(i|j)} \otimes
 \boldsymbol{c}'\cdot\hat{\boldsymbol{\Pi}}{}^{(k)}
+
 \tfrac12 \hat{\mathscr{B}}'{}^{(i|j)} \otimes
 \boldsymbol{c}\cdot\hat{\boldsymbol{\Pi}}{}^{(k)},
\end{equation}
where $\hat{\mathscr{B}}'{}^{(j|k)}(\omega)$ is the same as $\hat{\mathscr{B}}{}^{(j|k)}(\omega)$ defined in \eqref{eq.GKMR3} with the primes reversed, and $\boldsymbol{c}$ and $\boldsymbol{c}'$  are normalized vectors
(as well as  $\boldsymbol{a}$, $\boldsymbol{a}'$, $\boldsymbol{b}$, and $\boldsymbol{b}'$ involved in the definition of  $\hat{\mathscr{B}}{}^{(j|k)}$).
Expanding expression \eqref{eq.TP0} shows that $\hat{\mathscr{S}}{}^{(i|j|k)}$ is invariant upon a permutation of its indices, provided the names of the unit vectors $(\boldsymbol{a},\boldsymbol{b},\boldsymbol{c})$ and $(\boldsymbol{a}',\boldsymbol{b}',\boldsymbol{c}')$
undergo the same permutation. In the following we arbitrarily chose the order $(i,j,k)=(0,1,2)$.

 Similarly to what occurs for the two-mode operator, the principle of local realism, if correct, should predict that
a measure of the operator $\hat{\mathscr{S}}{}^{(0|1|2)}$ yields results $\pm 2$.  The system violates this principle
when the average of the operator \eqref{eq.TP0} is larger than two. This is often referred to as violation of 
Svetlichny inequality. It is important to note that the observable $\hat{\mathscr{S}}{}^{(0|1|2)}$ is
specially designed in order to be sensitive to {\it genuine tripartite nonlocality} (see discussions in Refs. \cite{Svetlichny1987,Collins2002,Bancal2011,Brunner2014} and references therein): a tripartite system can involve nonlocal correlations between any of its bipartitions and still not violate the Svetlichny inequality $\langle 
\hat{\mathscr{S}}{}^{(0|1|2)}\rangle <2$. A simple example of a system which displays two mode nonlocality but does not pass the Svetlichny test is presented in Appendix \ref{app:analogue} [see Eq. \eqref{B_3modes_fbasis} and the discussion below].

\begin{figure}
    \centering
    \includegraphics[width=\linewidth]{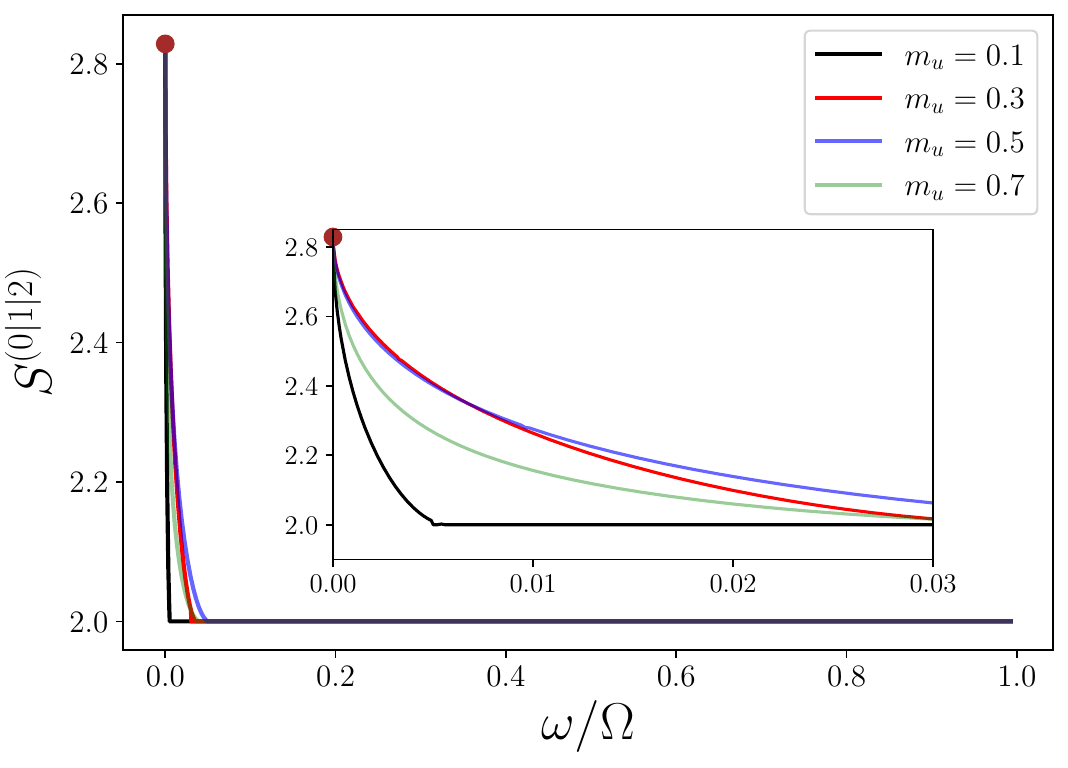}
    \caption{Zero temperature value of the tripartite parameter
      $S^{(0|1|2)}$ plotted as a function of energy for different
      realization of the waterfall configuration, each being
      characterized by the upstream Mach number $m_u$. The inset
      displays a blow up of the figure at low energy. The upper bound
      $2\sqrt{2}$ is marked by a brown dot. It is reached at
      $\omega=0$ in all configurations.}
    \label{fig.Trip0}
\end{figure}

\begin{figure}
\centering
\includegraphics[width=\linewidth]{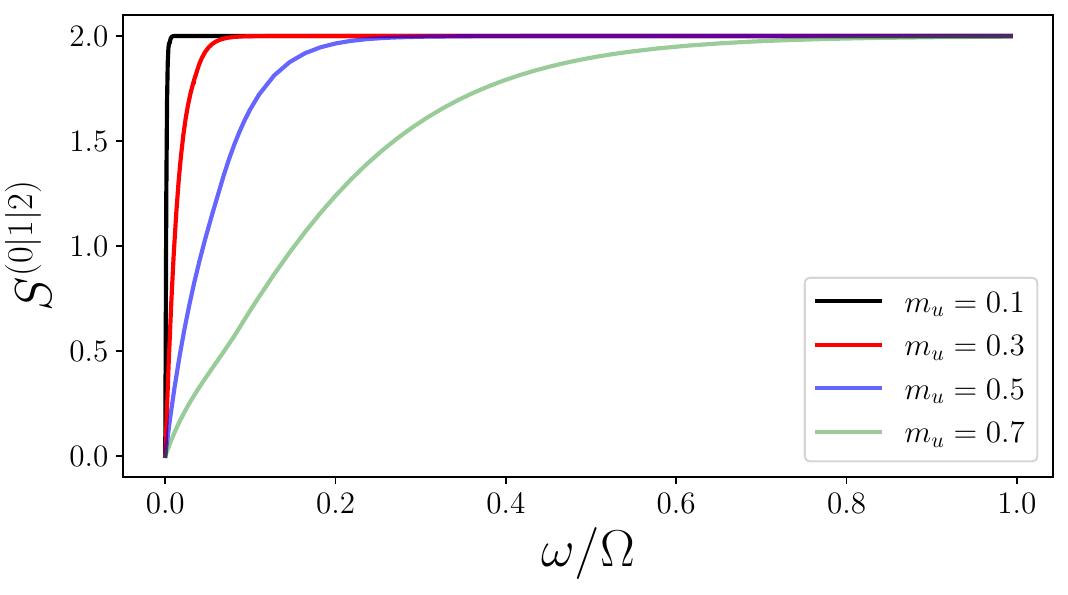}
\caption{Same as Fig. \ref{fig.Trip0} for a temperature
$T=0.05 \, gn_u$. At variance with the zero temperature
situation the tripartite measure \eqref{eq.TP7} is here always less than or equal to 2.}
\label{fig.Trip}
\end{figure}

For attempting to reach a maximum violation of Svetlichny (i.e., three-mode Bell) inequality it is necessary to choose a particular arrangement of the vectors $\boldsymbol{a}$, 
$\boldsymbol{a}'$, $\boldsymbol{b}$, $\boldsymbol{b}'$, $\boldsymbol{c}$ and  $\boldsymbol{c}'$ which maximizes the expectation value
 $\langle \hat{\mathscr{S}}{}^{(0|1|2)}\rangle$. 
Do do so, we resort to a genetic algorithm which is presented in Appendix \ref{app.maxi3} and numerically determines the quantity
\begin{equation}\label{eq.TP7}
S^{(0|1|2)}(\omega) \equiv \, \mbox{max}_{\boldsymbol{a},\boldsymbol{a}',\boldsymbol{b}, \boldsymbol{b}', \boldsymbol{c}, \boldsymbol{c}'}
\left\langle \hat{\mathscr{S}}{}^{(0|1|2)} (\omega) \right\rangle.
\end{equation}
It is shown in Appendix \ref{app.Maximization} 
that the tripartite parameter
$S^{(0|1|2)}(\omega)$ is bounded from above by $2\sqrt{2}$ 
[this is Eq. \eqref{eq.C3.5}]
and that this bound is reached at $\omega=0$ and $T=0$ [see Eq. \eqref{eq.B00max}].
The behavior of $S^{(0|1|2)}(\omega)$ at zero temperature is displayed in Fig. \ref{fig.Trip0} for different realizations of the waterfall configuration. Although the bound $2\sqrt{2}$ is reached in the long wave length limit for all the configurations, even a weak temperature is able to destroy the 
signal of nonlocality as illustrated in Fig. \ref{fig.Trip}.
This is connected to the loss of purity of the finite temperature system and
can be understood analytically, again in the long wave length limit, as discussed in Appendix \ref{app.maxi3}. This sensitivity to a small finite temperature probably precludes the experimental observation of tripartite nonlocality by means of the observable \eqref{eq.TP7}. However, although the zero temperature behavior is certainly difficult to observe, it is rich of fundamental insight on the nature of the state of the system, as we now discuss.

As argued in Appendix \ref{app.averages}, the fact that, at zero temperature and $\omega=0$ the system reaches the tripartite upper bound ($S^{(0|1|2)}=2\sqrt{2}$, see Fig. \ref{fig.Trip0}) mathematically stems from the fact that, under these conditions the system is in a pure state and displays perfect correlations in the following expectation values [see Eqs. \eqref{eq.lim2z}, \eqref{eq.lim2} and \eqref{eq.W9}]: 
\begin{equation}\label{eq.mean.0.0}
\begin{split}
\langle \hat{\Pi}_y^{(0)} \otimes \hat{\Pi}_y^{(1)}  \otimes \hat{\Pi}_z^{(2)} \rangle & = 
\langle \hat{\Pi}_y^{(0)} \otimes \hat{\Pi}_z^{(1)}  \otimes \hat{\Pi}_y^{(2)} \rangle \\ 
& = - \langle \hat{\Pi}_z^{(0)} \otimes \hat{\Pi}_y^{(1)}  \otimes 
\hat{\Pi}_y^{(2)} \rangle \\
& =
\langle \hat{\Pi}_z^{(0)} \otimes \hat{\Pi}_z^{(1)}  \otimes \hat{\Pi}_z^{(2)} \rangle = 1.
\end{split}
\end{equation}
All the other expectation values of products of three components of the $\hat{\boldsymbol{\Pi}}$ operator are zero. Since all the operators in \eqref{eq.mean.0.0} have only  $\pm 1$ as eigenvalues, this means that
each term  must reach its extremal value
and the vacuum mode of zero frequency 
$|0_{\omega=0}\rangle^{\rm in}$ must therefore be an eigenstate of the operators
appearing in \eqref{eq.mean.0.0}: 
\begin{subequations}\label{eq.VP}
\begin{align}
& \hat{\Pi}_y^{(0)} \otimes \hat{\Pi}_y^{(1)}  \otimes \hat{\Pi}_z^{(2)}
|0_{\omega=0}\rangle^{\rm in}
= + |0_{\omega=0}\rangle^{\rm in} , \label{eq.VPa} \\    
& \hat{\Pi}_y^{(0)} \otimes \hat{\Pi}_z^{(1)}  \otimes \hat{\Pi}_y^{(2)} 
|0_{\omega=0}\rangle^{\rm in}
= + |0_{\omega=0}\rangle^{\rm in} , \label{eq.VPb} \\
& \hat{\Pi}_z^{(0)} \otimes \hat{\Pi}_y^{(1)}  \otimes \hat{\Pi}_y^{(2)}
|0_{\omega=0}\rangle^{\rm in}
= - |0_{\omega=0}\rangle^{\rm in} , \label{eq.VPc} \\
& \hat{\Pi}_z^{(0)} \otimes \hat{\Pi}_z^{(1)}  \otimes \hat{\Pi}_z^{(2)}
|0_{\omega=0}\rangle^{\rm in}
= + |0_{\omega=0}\rangle^{\rm in} . \label{eq.VPd}
\end{align}
\end{subequations}
Relations \eqref{eq.VP}
define a state which exhibits the GHZ paradox, contradicting local hidden variable theories by means of a single measurement; cf. e.g., the discussion in \cite{Mermin1990a}. In the same line, the above noted fact that our system saturates the upper bound  $2\sqrt{2}$ of the Svetlichny parameter is a characteristic also shared by GHZ states. 
Another resemblance lies in the fact that the GHZ states 
possess maximal tripartite
entanglement, quantified by the residual tangle \cite{Coffman2000,Dur2000} while similarly, in the domain of temperature and energy where Eqs. \eqref{eq.VP} hold, the residual contangle \cite{Adesso2006a} which measures genuine tripartite entanglement in our {\it continuous variable} system diverges \cite{Isoard2021}.
Despite these clear similarities, there is a significant difference between the state
$|0_{\omega=0}\rangle^{\rm in}$ and the GHZ states commonly considered in quantum information theory: they have different behaviors upon partial tracing.
As is well known, taking the partial trace over one of the three modes of a GHZ state yields an unentangled mixed state.
At variance, partial tracing
the state $|0_{\omega=0}\rangle^{\rm in}$ over modes 0 or 1 leads to an entangled two-mode mixed state, 
as discussed in Sec. \ref{sec.Bell2} (see, e.g., the values $\Lambda^{(0|2)}(\omega=0)$
and $\Lambda^{(1|2)}(\omega=0)$ in Fig. \ref{fig3.1}). As we now argue, the explanation of this conundrum lies in the fact that, whereas the 
GHZ states are usually build on qubits, the state we consider is an 
infinite sum of degenerate GHZ states of a continuous variable system. 

To tackle this issue it is convenient to expand the state  
$|0_{\omega=0}\rangle^{\rm in}$ over the eigenstates of the operators $\hat{\Pi}_x^{(j)}$ ($j=0$, 1 or 2). As discussed in Appendix \ref{app.pseudo.spin} these eigenstates 
can be 
written as $|x^{\pm}_{n}\rangle_{j}$: they are
labeled by their eigenvalue ($\pm 1$) plus another integer index ($n$ in the above expression) associated to the infinite degeneracy of both eigenvalues. 
Indeed, at variance with what occurs for a regular spin operator, the projection of the pseudo-spin \eqref{eq.GKMR1}
over a given axis (here $\hat{\Pi}_x$) has infinitely degenerate eigenvalues, i.e.,  
there exists an infinite number of mutually orthogonal eigenstates with the same eigenvalue ($+1$ or $-1$):
\begin{equation}
\forall n  \in\mathbb{N}, \quad   \hat{\Pi}_x^{(j)} |x^{\pm}_{n}\rangle_{j}
    = \pm |x^{\pm}_{n}\rangle_{j},
\end{equation}  
whereas
\begin{equation}
\forall (n,m)  \in\mathbb{N}^2, \quad  _j\langle x^{\pm}_{n}|x^{\pm}_{m}\rangle_{j}
=\delta_{n,m}.
\end{equation}
This can be shown by directly constructing the eigenstates of $\hat{\Pi}_x$ from the number states, see Eqs. \eqref{eq.ps4a} and \eqref{eq.ps4}.
The expansion of $|0_{\omega=0}\rangle^{\rm in}$  over the complete basis formed by these states is of the form:
\begin{equation}\label{eq.devel}
|0_{\omega=0}\rangle^{\rm in}
=\sum_{\substack{\sigma_0,\sigma_1,\sigma_2\\l,m,n}}
C^{\sigma_0,\sigma_1,\sigma_2}_{l,m,n}
|x^{\sigma_0}_{l},x^{\sigma_1}_{m},x^{\sigma_2}_{n}
\rangle.
\end{equation}
In the summation appearing in the above expression, $\sigma_0$, $\sigma_1$ and $\sigma_2=\pm$ whereas $(l,m,n)\in \mathbb{N}^3$ and we dropped the indices $j=0$, 1 or 2 of the kets for legibility. 

It follows from relations
\eqref{eq.VP} and
\eqref{eq.ps7} that:
\begin{equation}
\begin{split}
    C_{l,m,n}^{-\sigma_0,-\sigma_1,-\sigma2}
    =-\sigma_0\sigma_1 \,& C_{l,m,n}^{-\sigma_0,-\sigma_1,-\sigma_2}\\
    =-\sigma_0\sigma_2 \,& C_{l,m,n}^{-\sigma_0,-\sigma_1,-\sigma_2}\\
    =\sigma_1\sigma_2 \,& C_{l,m,n}^{-\sigma_0,-\sigma_1,-\sigma_2}. 
\end{split}
\end{equation}
This imposes that the only nonzero coefficients in expansion \eqref{eq.devel} are those for which $\sigma_1=\sigma_2=-\sigma_0$. For such coefficients Eqs. \eqref{eq.VPd} and \eqref{eq.ps7} impose that
\begin{equation}
    C^{+--}_{l,m,n}=C^{-++}_{l,m,n}
    \equiv C_{l,m,n}.
\end{equation}
Expansion \eqref{eq.devel} thus simplifies to
\begin{equation}\label{eq.devel2}
\begin{split}
|0_{\omega=0}\rangle^{\rm in} =\sum_{l,m,n}
C_{l,m,n}
\Big( & 
|x^{+}_{l},x^{-}_{m},x^{-}_{n}\rangle\,
+\\
&
|x^{-}_{l},x^{+}_{m},x^{+}_{n}\rangle
\Big),
\end{split}
\end{equation}
which shows that the vacuum $|0_{\omega=0}\rangle^{\rm in}$
of the $\hat{b}_j(\omega=0)$ operators (the ingoing ground state) is an 
infinite sum of degenerate GHZ states. It is this property which enables the reduced state obtained after partial tracing over one 
mode to remain entangled despite the GHZ nature of the system. Indeed, tracing over mode 
0 for instance leads to a reduced density matrix
\begin{equation}\label{eq.rho0a}
\begin{split}
\mbox{Tr}_{(0)}\Big( 
|0_{\omega=0}\rangle^{\rm in} \; ^{\rm in}\langle 0_{\omega=0}|
    \Big)& =\\
    \sum_{m,n,\mu,\nu} 
{\cal C}^{(0)}_{m,n,\mu,\nu} & \Big( 
|x^{-}_{m} x^{-}_{n}\rangle \, \langle x^{-}_{\mu} x^{-}_{\nu} | \\
&  +
|x^{+}_{m} x^{+}_{n}\rangle \, \langle x^{+}_{\mu} x^{+}_{\nu} | \Big).
\end{split}
\end{equation}
In this expression the kets and bras concern modes 1 and 2 (mode 0 has been traced out) and
\begin{equation}\label{rho0b}
    {\cal C}^{(0)}_{m,n,\mu,\nu}=\sum_{l=0}^{\infty} C^*_{l,\mu,\nu} C_{l,m,n}.
\end{equation}
If the eigenvalues $+1$ and $-1$ of operator $\hat{\Pi}_x$ were non-degenerate, this would impose $m=\mu$ and $n=\nu$ in expression \eqref{eq.rho0a} and the corresponding reduced state would be clearly separable. Nothing similar occurs in our situation, and indeed the reduced states are typically entangled, as shown  in Sec. \ref{sec.Bell2}.

Since we now understand the exact GHZ nature of the zero temperature and $\omega=0$ state of the system, it is of interest to quantify to what extent this feature persists at finite 
temperature and finite energy. To this aim, we use the genetic algorithm presented in Appendix \ref{app.maxi3} to
compute the optimum of the Mermin parameter \cite{Mermin1990,Klyshko1993}
\begin{equation}\label{eq.M1}
M^{(0|1|2)}(\omega) \equiv \, \mbox{max}_{\boldsymbol{a},\boldsymbol{a}',\boldsymbol{b}, \boldsymbol{b}', \boldsymbol{c}, \boldsymbol{c}'}
\left|\left\langle \hat{\mathscr{M}}{}^{(0|1|2)} (\omega) \right\rangle\right|,
\end{equation}
where
\begin{equation}\label{eq.M2}
\begin{split}
\hat{\mathscr{M}}^{(0|1|2)}(\omega)= & 
-\boldsymbol{a}\cdot\hat{\boldsymbol{\Pi}}^{(0)}\otimes
\boldsymbol{b}\cdot\hat{\boldsymbol{\Pi}}^{(1)}\otimes
\boldsymbol{c}\cdot\hat{\boldsymbol{\Pi}}^{(2)} \\
&+\boldsymbol{a}\cdot\hat{\boldsymbol{\Pi}}^{(0)}\otimes
\boldsymbol{b}'\cdot\hat{\boldsymbol{\Pi}}^{(1)}\otimes
\boldsymbol{c}'\cdot\hat{\boldsymbol{\Pi}}^{(2)} \\
&+\boldsymbol{a}'\cdot\hat{\boldsymbol{\Pi}}^{(0)}\otimes
\boldsymbol{b}\cdot\hat{\boldsymbol{\Pi}}^{(1)}\otimes
\boldsymbol{c}'\cdot\hat{\boldsymbol{\Pi}}^{(2)} \\
&+\boldsymbol{a}'\cdot\hat{\boldsymbol{\Pi}}^{(0)}\otimes
\boldsymbol{b}'\cdot\hat{\boldsymbol{\Pi}}^{(1)}\otimes
\boldsymbol{c}\cdot\hat{\boldsymbol{\Pi}}^{(2)}.
\end{split}
\end{equation}
We note here that $\hat{\mathscr{S}}^{(0|1|2)}=\tfrac12 \hat{\mathscr{M}}^{(0|1|2)} + \tfrac12 \hat{\mathscr{M}}'{}^{(0|1|2)}$, where $\hat{\mathscr{M}}'{}^{(0|1|2)}$ is the same as
$\hat{\mathscr{M}}^{(0|1|2)}$ with the prime reversed \cite{Collins2002}.

\begin{figure}
    \centering
    \includegraphics[width=\linewidth]{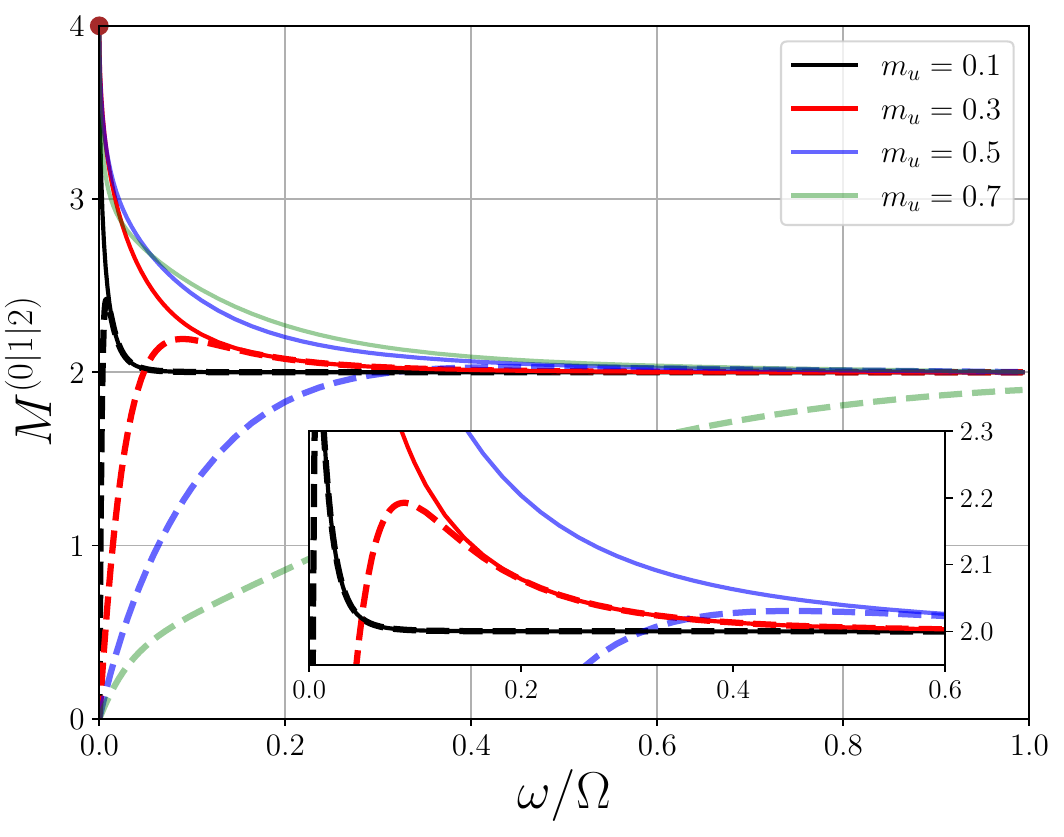}
    \caption{Mermin parameter $M^{(0|1|2)}(\omega)$ plotted for
      different waterfall configurations. The continuous lines
      starting from $M^{(0|1|2)}=4$ at $\omega=0$ are zero temperature
      results and the thick dashed ones (with $M^{(0|1|2)}(0)=0$)
      correspond to $T=0.1\, g n_u$. The inset is a blow up of the
      figure around the region $M^{(0|1|2)}=2$ for
      $\omega/\Omega\le 0.6$.}
    \label{fig.Mermin_WF}
\end{figure}

The largest possible value of the Mermin parameter \eqref{eq.M1} is 4. This upper bound is reached for a state verifying the relations \eqref{eq.mean.0.0}, such as a GHZ state or the low wavelength 
component of the ground state of our system as just 
discussed\footnote{From \eqref{eq.mean.0.0} it is clear that at zero temperature and $\omega=0$ the maximum \eqref{eq.M1} is obtained when $\boldsymbol{a}= \boldsymbol{b}'= \boldsymbol{c}'= \boldsymbol{e}_z$ and $\boldsymbol{a}'= \boldsymbol{b}= \boldsymbol{c}= \boldsymbol{e}_y$
and reaches the upper bound 4.}.  Indeed, at $T=0$, $M^{(0|1|2)}(0)=4$ for all black hole configurations. This is illustrated in Fig. \ref{fig.Mermin_WF}
in which $M^{(0|1|2)}$ is plotted as a function of $\omega$ for different values of $m_u$, at zero and finite temperature ($T=0$ and $T=0.1\, gn_u$). 
If the departure of $M^{(0|1|2)}$ from its upper bound is taken as an indication of how much the system differs from a GHZ state, this suggests that the GHZ character of the state is restricted to the low energy and low temperature sector.
We stress however that such a criterion is only indicative: The  Svetlichny and Mermin parameters provide useful bounds, but not proper measures. This is illustrated by their poor effectiveness for evaluating genuine tripartite entanglement: 
It is known \cite{Collins2002} that if
\begin{subequations}\label{eq.M3}
\begin{align}
    M^{(0|1|2)}& >2\sqrt{2},\label{eq.M3a} \\ 
    \mbox{or}\quad
S^{(0|1|2)}& >2,\label{eq.M3b}
\end{align}
\end{subequations}
the system exhibits genuine three-mode entanglement. Figs. \ref{fig.Trip0} and \ref{fig.Mermin_WF} show that this type of entanglement is certainly reached at $T=0$ for low energy. However, the criteria \eqref{eq.M3} are here too restrictive, since the computation of the residual contangle done in Ref. \cite{Isoard2021} demonstrates that at $T=0$ the system is genuinely tripartite entangled for {\it all} energies and {\it all} configurations. This general conclusion certainly cannot be reached by 
inspecting in Figs. \ref{fig.Trip0} and \ref{fig.Mermin_WF} in which domain of energy and for which value of $m_u$ the criteria \eqref{eq.M3} are met.
However, these criteria are valid and can be useful, for instance at finite temperature. In this case the state of the system is not pure, the evaluation of the residual contangle appears to be very difficult, and the criterion \eqref{eq.M3a} is the only way we know how to demonstrate genuine tripartite entanglement, which, as can be inferred from the tendency displayed in Fig. \ref{fig.Mermin_WF}, is reached at low $T$, low $\omega$ and small values of $m_u$.

Finally, the Mermin parameter \eqref{eq.M1} is also an interesting witness of nonlocality.
A local hidden variable theory predicts that it  should
verify the Mermin-Klyshko inequality $M^{(0|1|2)}\le 2$.
It can be seen from Fig. \ref{fig.Mermin_WF} that this inequality is violated at zero temperature for all waterfall configurations. For $T>0$ instead,
$M^{(0|1|2)}(0)=0$.
The Mermin parameter has the same behaviour as the Svetlichny parameter, for the same reason: At finite temperature the state of the system is no longer pure, and in this case all the expectation values of products of three components of the pseudo-spin tend to zero in the long wavelength limit, see discussion in Appendix \ref{app.averages}.
However, contrarily to what occurs for the Svetlichny parameter, there are accessible finite temperature situations where the Mermin parameter is larger that the nonlocality threshold $M^{(0|1|2)}=2$ (compare Figs. \ref{fig.Mermin_WF} and \ref{fig.Trip}). In that respect, the Mermin parameter may even reveal more useful that the  
CHSH parameter \eqref{eq.GKMR6}. For instance, at $T=0.1 g n_u$, for $m_u=0.3$ the largest value of $M^{(0|1|2)}$ is 2.19 (as seen from Fig. \ref{fig.Mermin_WF}), higher than the largest values reached by both CHSH parameters $B^{(0|2)}$ and $B^{(0|1)}$ in the same situation (2 and 2.017, respectively).

\section{Conclusion}\label{sec.conclusion}

In this work we conducted a systematic theoretical  study of violation of bipartite and tripartite Bell inequalities in an analogue of black hole realized in the flow 
of a quasi-one dimensional Bose-Einstein condensate. There is a reasonable hope to witness bipartite entanglement in such systems, and also possibly bipartite nonlocality whereas the observation of 
signatures of genuine tripartite nonlocality would presumably be more difficult. The reason is that violation of Svetlichny inequality does not resist much to an 
increase in temperature. In that respect it is worth underlying that, from the three black hole configurations we have considered, the waterfall configuration is the one for which the signature of nonlocality is the less sensitive to an increased temperature: compare for instance the finite temperature signal of Fig. \ref{fig.Mermin_WF} with the one of Figs. \ref{fig.Mermin_DP} and \ref{fig.Mermin_FP}.

In an analog system, the essential ingredient for observing Hawking radiation is roughly the same as the one initially invoked by Hawking in the gravitational context \cite{Hawking1975}: the vacuum of the outgoing modes is not the same as the vacuum of the ingoing modes.
In the BEC system we consider, this is physically due to the fact that the asymptotic upstream region is subsonic while the asymptotic downstream region is supersonic. A flow of this type induces the mismatch of the asymptotic dispersion relations illustrated in Fig. \ref{fig2}, with additional (negative norm) channels in the downstream region, resulting in the Bogoliubov transformation \eqref{eq.excit3} from which stems relation \eqref{eq.entang1} between the ingoing and outgoing vacua. The other important physical ingredient is the sonic character of the dispersion relation in the long wavelength limit. This results in
the low energy divergence of the coefficients of the $S$ matrix involving the incoming negative norm mode (the $S_{j,2}$ coefficients with our notations) \cite{Robertson2012} and, {\it in fine}, in the finite Hawking temperature \cite{Macher2009,Recati2009}. This being said, the Bogoliubov transform we consider is not an exotic one; it is attached to a standard quadratic Bose Hamiltonian \cite{Blaizot1986} and to the production of correlated pairs of quasi-particles. This is the reason why, as illustrated in Fig. \ref{fig.optical.model}, we can mimic our system by a simple optical setup involving a standard parametric downconvertion process.
From this perspective the existence of three and not simply two modes (which might be considered as atypical in the Bogoliubov context) simply stems from the presence of a beam-splitter in the optical setting. Going back to the gravitational point of view, it has been argued in \cite{Isoard2021} that this beam splitter embodies the scattering of Hawking radiation by the geometry of the black hole, i.e., it effectively reproduces the greybody factor. The fact that such a simple pair production process induces {\it genuine} tripartite entanglement and {\it genuine} tripartite non-locality is far from being intuitive and is an important results of Ref. \cite{Isoard2021} and of the present study.

It results from our Bogoliubov treatment that we describe the system as a Gaussian state. This is certainly a sensible first order hypothesis, but recent studies \cite{Robertson2018,Chatrchyan2021,Butera2023,deNova2023} showed that nonlinear effects might significantly affect quantum emission processes in the context of analogue physics. It would then be of great interest to evaluate in detail the amplitude and the relevance of nonlinear back reaction effects on acoustic Hawking radiation in the black hole analogues we consider.

Within the Gaussianity assumption, the main theoretical tool of our study is the covariance matrix. As indicated in Sec. \ref{sec.model}, indirect techniques have been used for measuring some of the entries of this matrix in BEC physics. Other systems, such as exciton-polaritons in microcavities, or more generally setups involving nonlinear light, could enable other types of measurements, with possibly more direct access to the phases of the averaged quadratures. 
Our approach can be adapted with little modification to such systems. For the quantities we are interested in, a key element is the long wavelength behavior of the dispersion relation (and of the corresponding $S$ matrix). For instance the GHZ nature of the long wavelength modes is generic in any system 
with a sonic-like low energy dispersion relation. The observables studied in the text might however be modified in a system with a different long wave-length behavior, for instance in the presence of massive modes, as is the case in the polariton context, but also in coherently coupled two-component BECs.

A fundamental outcome of our work 
stems from a flaw in the analogy. Our analogue system breaks Lorentz invariance, but
paradoxically this apparent shortcoming (characteristic of 
analogue physics \cite{Jacobson1991,Unruh1995,Corley1996,Corley1998}) turns into an advantage: It opens the possibility of tripartite entanglement and nonlocality, thanks to
the existence not only of a Hawking and a partner mode, but also of a third mode we denote as the companion
This peculiar nonlocal tripartite configuration,
together with
the continuous nature of the degrees of freedom, 
induces a surprising property of the system: the
long wavelength quantum modes consist in a superposition
of  degenerate GHZ
states which,  at variance with GHZ states build on
qubits, remains entangled after partial tracing.
We expect this feature to be generic, thus suggesting that condensed matter analogues may indeed open new
prospects of robust information protocols and provide efficient platforms to study
the flow of multipartite quantum information.

\begin{acknowledgments}
  We thank M. J. Jacquet, J. Martin and M. Walschaers for inspiring
  discussions.  We also gladly acknowledge C. E. Lopetegui for
  relevant remarks concerning continuous-variable systems. We are
  particularly grateful to A. Buchleitner for thorough exchanges on
  many of the aspects addressed in this work.  This project has
  received support by the DFG funded Research Training Group ``Dynamics
  of Controlled Atomic and Molecular Systems'' (RTG 2717).
\end{acknowledgments}

\appendix

\section{Different black hole configurations}\label{app.different}

In this Appendix we first briefly present the range of validity of expansion \eqref{eq.model2} and then introduce some stationary solutions of the classical background field $\Phi(x)$ corresponding to an analog black hole.

For a quasi-1D guided BEC transversely trapped by a harmonic potential with angular frequency $\omega_\perp$, expansion \eqref{eq.model2} is valid in the ``1D mean field regime''
\cite{Menotti2002} defined by
\begin{equation}\label{Appa.1}
    \left(\frac{a}{a_\perp}\right)^2\ll n_{\rm typ} a \ll 1,
\end{equation}
where $a$ is the 3D s-wave scattering length,
$a_\perp=\sqrt{\hbar/m\omega_\perp}$ and $n_{\rm typ}$ is the typical linear density (number of atoms per unit length). The left inequality in \eqref{Appa.1} ensures that the finite phase coherence length (induced by long wavelength quantum fluctuations) is exponentially large compared to the healing length. The inequality at the riht ensures that the transverse degrees of freedom are frozen. For a
transverse trap of frequency of 1 kHz, one gets
$(a_\perp/a)^2=1.7\times 10^{-5}$ and
$2.6\times 10^{-4}$, for $^{23}$Na and $^{87}$Rb, respectively. Hence the domain
of validity of the 1D mean field approximation used in the present
work typically ranges over four orders of magnitudes in density.

In the regime where \eqref{Appa.1} holds, the background classical field $\Phi$ of Eq. \eqref{eq.model2} is solution of a Gross-Pitaekskii equation which is the stationary and classical version of Eq. \eqref{eq.model1}: 
\begin{equation}\label{eq.model1bis}
    -\frac{\hbar^2}{2m}\partial_x^2\Phi
    +[U(x) + g \, n-\mu]\Phi =0 ,
\end{equation}
where $n=|\Phi|^2$ and $g=2\hbar\omega_\perp a$. An analogue black hole horizon is realized if the flow is subsonic far upstream (for $x\to -\infty$ in our convention) and takes the form of a downstream supersonic plane wave (for $x>0$). The corresponding classical field behaves as
\begin{equation}\label{eq.model3}
\Phi(x)=\left\{\begin{array}{lcl}
\sqrt{n_u}\exp({\rm i} k_u x)\,\phi_u(x) & \mbox{for} & x\le 0, \\
\sqrt{n_d}\exp({\rm i} k_d x)\exp({\rm i} \beta_d) & \mbox{for} & x\ge 0.
\end{array}\right.
\end{equation}
In this expression $\lim_{x\to-\infty} \phi_u(x)=\exp({\rm i}
\beta_u)$, where $\beta_u$ is a constant, as is $\beta_d$. In all the text the indices $u$ and $d$ refer to the upstream and downstream flow. For instance
$n_u$ and $n_d$ are the upstream and downstream asymptotic densities. Also
$k_\alpha = m V_\alpha/\hbar$ ($\alpha=u$ or $d$), where $V_u$ ($V_d$) is
the asymptotic upstream (downstream) flow velocity. The asymptotic velocities of sound
$c_u$ and $c_d$ are defined by $m c_\alpha^2=g_\alpha n_\alpha$, where\footnote{We consider a possible
position-dependent nonlinear coefficient for being able to treat the flat profile
configuration of Ref. \cite{Carusotto2008}.}
$g_{u,d}=\lim_{x\to-\infty,+\infty}g(x)$. The
healing lengths and Mach numbers are defined as $\xi_\alpha=\hbar/(m c_\alpha)$ and
$m_\alpha=V_\alpha/c_\alpha$, respectively.
The form of the flow pattern is specified by the value of the above parameters and 
by the function $\phi_u(x)$. In the remaining of this subsection we briefly describe the 3 configurations studied in the present work and refer to \cite{Larre2012} for a detailed presentation 
of these configurations
with the precise values of the corresponding parameters.

The first configuration we consider is an idealized one, introduced in
Ref. \cite{Carusotto2008} which we denote as ``flat profile''. It
consists in a constant uniform plane wave flow with $\phi_u(x)=1$,
$\beta_d=0$, $n_u=n_d\equiv n_0$ and $k_u=k_d\equiv k_0$. Such a
configuration can realize an analog black hole in the presence of a
piece-wise nonlinear parameter $g(x)=g_u\Theta(-x)+g_d\Theta(x)$,
supplemented by an also piece-wise potential $U(x)$, as explained in
Refs. \cite{Carusotto2008,Larre2012}. In the two other configurations
we consider $g$ is a constant.

The second configuration is denoted as ``waterfall'': The potential
$U(x)$ is a step function and the upstream profile $\phi_u(x)$ is half that of a dark
soliton. It is close to the experimental realization of the Technion
group \cite{De_nova_2019,Kolobov_2021} and catches important aspects
of the density-correlation pattern reported in \cite{De_nova_2019},
with nonetheless some caveats, see \cite{Isoard2020}.

The third and last configuration is denoted as ``delta peak'': the
potential $U(x)$ is a repulsive Dirac distribution located at $x=0$
and the upstream profile $\phi_u(x)$ is a part of a dark soliton, but not exactly
one half of it, as is the case in the waterfall configuration.
\begin{figure}
\centering
\includegraphics[width=\linewidth]{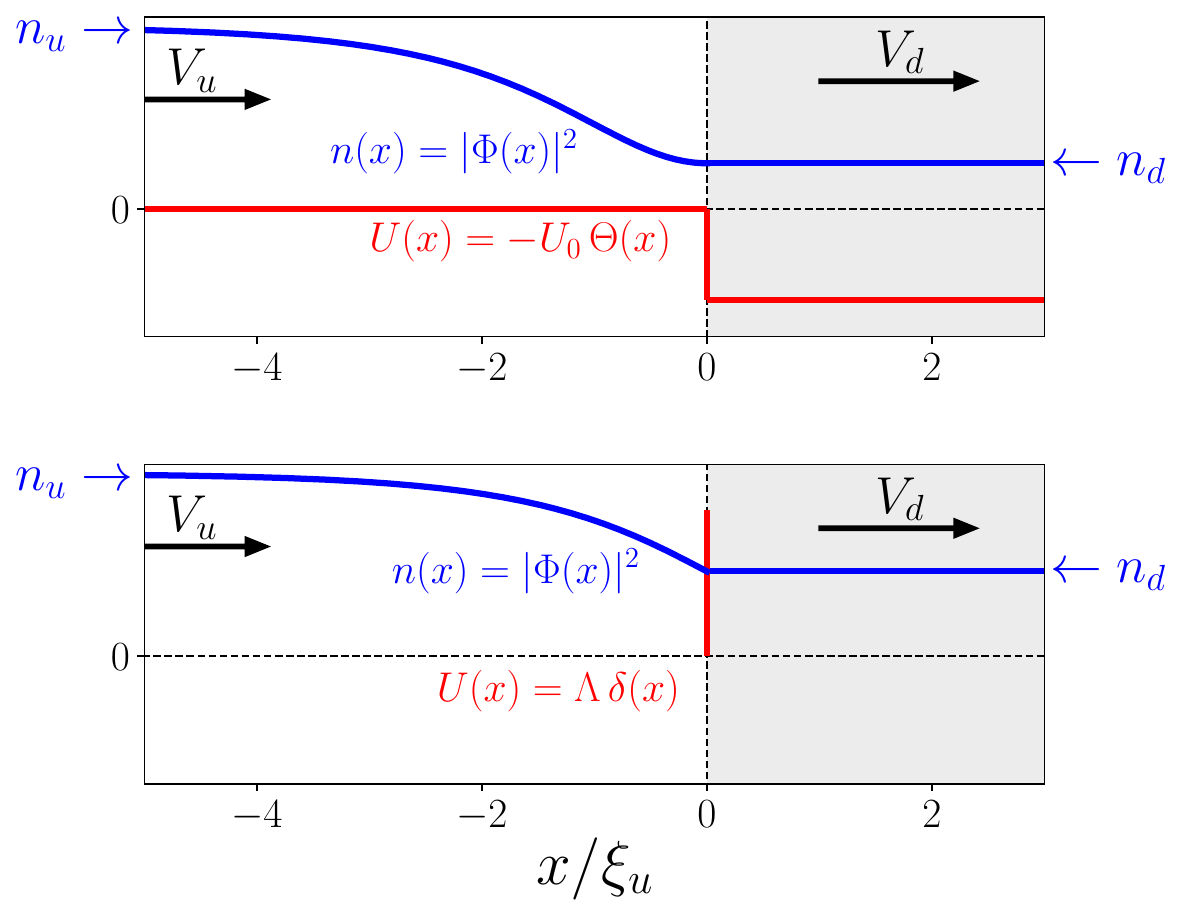}
\caption{Schematic representation of the background density profile of
  the waterfall configuration (upper plot) and of the delta peak
  configuration (lower plot). The shaded region corresponds to the
  interior of the analog black hole (see the main text). The whole
  $x>0$ region is supersonic, while the upstream region is
  asymptotically subsonic (i.e., in the limit
  $x\to -\infty$).}\label{fig1}
\end{figure}
Both configurations are depicted in Fig. \ref{fig1}. In this figure the region $x>0$ is shaded in order to remind that it corresponds to the interior of the analog black hole. It is however important to recall that the precise location of the horizon separating the interior and the exterior of an analog black hole is ill-defined, see, e.g., the discussion in Sec. II.A of Ref. \cite{Isoard2021}.

Note that, while the upstream and downstream Mach numbers can be fixed independently in the flat profile configuration (with the only constrain that $m_u<1<m_d$), these two quantities are not independent in the waterfall and delta peak configurations: 
\begin{subequations}\label{eq.model4}
\begin{align}
m_d=m_u^{-2} \quad & \mbox{for waterfall}, \label{eq.model4a} \\
\frac{m_d}{m_u}= \left( 
\frac{\displaystyle -1+\sqrt{1+8 m_u^{-2}}}{\displaystyle 2}\right)^{3/2} 
\; &
\mbox{for delta peak.} \label{eq.model4b}
\end{align}
\end{subequations}

\section{Covariance matrices}\label{app.sigma}

The $6\times 6$ covariante matrix $\sigma$ defined in Eq. \eqref{eq.entang4} can be
written in terms of $2\times 2$ submatrices $\sigma_i$ and
$\varepsilon_{ij}$:
\begin{equation}\label{eq.apsig1}
    \sigma=\begin{pmatrix} 
    \sigma_0 & \varepsilon_{01} & \varepsilon_{02} \\
    \varepsilon_{01}^{\rm \sss T } & \sigma_1 & \varepsilon_{12} \\
    \varepsilon_{02}^{\rm \sss T } & \varepsilon_{12}^{\rm \sss T } & \sigma_2
    \end{pmatrix},
\end{equation}
where
\begin{equation}\label{eq.apsig2}
\sigma_{i} =
\begin{pmatrix}
\left\langle  2\hat{q}_{i}^2 \right\rangle 
& \left\langle  \hat{q}_{i} \, \hat{p}_{i} + \hat{p}_{i} \, \hat{q}_{i} \right\rangle  \\
\left\langle    \hat{q}_{i}\, \hat{p}_{i} + \hat{p}_{i} \, \hat{q}_{i}\right\rangle 
& \left\langle 2\hat{p}_{i}^2 \right\rangle \\
\end{pmatrix}
,
\end{equation}
and
\begin{equation} \label{eq.apsig3}
\varepsilon_{ij} =
2 \, 
\begin{pmatrix}
 \left\langle  \hat{q}_{i} \, \hat{q}_{j} \right\rangle 
& \left\langle   \hat{q}_{i} \, \hat{p}_{j} \right\rangle   \\
 \left\langle \hat{p}_{i} \, \hat{q}_{j}  \right\rangle  
&  \left\langle   \hat{p}_{i} \, \hat{p}_{j}  \right\rangle  \\
\end{pmatrix}.
\end{equation}
In the case we consider, the 
definitions \eqref{eq:qetp} and the
specific form of the transformation \eqref{eq.excit3}
result in the following expressions:
\begin{equation} \label{eq.apsig2bis}
\sigma_{i} =\left(1 + 2\,\langle \hat{c}_i^\dagger \hat{c}_i \rangle \right)\,
\mathds{1}_{2},
\end{equation}
\begin{equation}\label{eq.apsig4}
\varepsilon_{01} =2
\begin{pmatrix}
  \text{Re}  \langle \hat{c}_0 \, \hat{c}_1^\dagger \rangle  &
  -\text{Im}  \langle \hat{c}_0 \, \hat{c}_1^\dagger \rangle  \\
  \text{Im}   \langle \hat{c}_0 \, \hat{c}_1^\dagger \rangle  &
  \text{Re}  \langle \hat{c}_0 \, \hat{c}_1^\dagger \rangle 
\end{pmatrix},
\end{equation}
and (for $i=0$ or 1)
\begin{equation}\label{eq.apsig5}
\varepsilon_{i2}  = 2
\begin{pmatrix}
  \text{Re}  \left\langle \hat{c}_i \, \hat{c}_2 \right\rangle  &
  \text{Im}  \left\langle \hat{c}_i \, \hat{c}_2 \right\rangle  \\
  \text{Im}   \left\langle \hat{c}_i \, \hat{c}_2 \right\rangle  &
  -\text{Re}  \left\langle \hat{c}_i \, \hat{c}_2 \right\rangle 
\end{pmatrix}.
\end{equation}
In the limit of zero temperature the system is in a three mode pure
Gaussian state. Its covariance matrix \eqref{eq.apsig1} can accordingly be
brought by LLUBOs (local linear unitary Bogoliubov transformations) to
a ``standard form'' in which matrices $\sigma_{i}$ are proportional to
the identity [as they already are, cf. \eqref{eq.apsig2bis}] and matrices $\varepsilon_{ij}$ are
diagonal \cite{Duan2000,Adesso2006a}.  After this operation
the matrices $\varepsilon_{ij}$ take the following form \cite{Isoard2021}
\begin{equation}
\label{eq.apsig6}
\varepsilon_{01}
= 2 \,|\langle \hat{c}_0 \, \hat{c}_1^\dagger \rangle|  \mathds{1}_2\; , 
\;\;
\varepsilon_{i2} = 2 \, |\langle \hat{c}_i \, \hat{c}_2 \rangle| \, \sigma_z,
\end{equation}
where $i\in\{0,1\}$ and $\sigma_z$ is the third Pauli matrix.

The situation at finite temperature is less simple. The system is in a
mixed state with no special symmetry, and the covariance matrix
\eqref{eq.apsig1} cannot be put in a standard form where the matrices
$\varepsilon_{ij}$ are all diagonal \cite{Adesso2006a}. However, the situation simplifies again if
one is interested in bipartite entanglement only (say, between modes $i$
and $j$). In this case one should trace out the third mode (let's denote it by $k$)
which simply amounts to remove from the total covariance matrix \eqref{eq.apsig1} the
two rows and two columns where index $k$ appears. The remaining
$4 \times 4$ covariance matrix associated with the reduced two-mode
state reads
\begin{equation}\label{eq.apsig7}
\sigma^{(i|j)}=
\begin{pmatrix}
\sigma_i & \varepsilon_{ij} \\
\varepsilon_{ij}^{\rm \sss T } & \sigma_{j} 
\end{pmatrix},
\end{equation}
where the $2\times 2$ blocks are the same as the ones in \eqref{eq.apsig1}.
This reduced covariance matrix
can always be brought by LLUBOs to its standard form
\cite{Duan2000}, in which the matrix $\varepsilon_{ij}$ takes again
the form \eqref{eq.apsig6} with here an average taken over the finite
temperature state, as explained in the main text, Sec. \ref{subsec.DM}.

It has been argued in \cite{Isoard2021} that an efficient measure of
bipartite entanglement was given by the ``PPT measure'' $\Lambda^{(i|j)}\equiv 1-\nu^-_{(i|j)}$
where $\nu^-_{(i|j)}$   is the lowest simplectic eigenvalue of the partial
transpose of $\sigma^{(i|j)}$ \footnote{The partial transposition
corresponds to a mirror reflection in phase space which inverts the $p_j$ coordinate, leaving $q_i$, $p_i$ and $q_j$ unchanged \cite{Simon2000}. The resulting covariance matrix can be brought by means of a symplectic transform to a diagonal form \cite{williamson1936}. The corresponding diagonal elements are the simplectic eigenvalues. They are twice degenerate, and in our $4\times 4$ case there are thus two such eigenvalues: $\nu^-_{(i|j)}$ and $\nu^+_{(i|j)}$, with $\nu^\pm_{(i|j)}\in \mathbb{R}^+$ and $\nu^-_{(i|j)} \le \nu^+_{(ij)}$.}. The largest entanglement corresponds to $\Lambda^{(i|j)}=1$ while separability is reached 
when $\Lambda^{(i|j)}<0$. This separability condition can be shown to be equivalent to the Peres-Horodecki criterion 
\cite{Peres1996,Horodecki1996}. Contrarily to other observables, 
such as the Cauchy-Schwarz criterion  or the generalized Peres-Horodecki parameter which have been often used in the domain, the PPT measure has the advantage of being an entanglement monotone. 
Other observables have been used in the context of analog gravity, which are monotonous measures of entanglement, such as the 
entanglement entropy, the entanglement of formation or the logarithmic negativity, but they all have some drawbacks: the state 
our our system is mixed in a finite temperature situation, which discards the entanglement entropy as a possible measure. The 
entanglement entropy generalizes for mixed states to the entanglement of formation \cite{Bennet1996}, but this quantity is 
not easily determined in non symmetric two-mode Gaussian states such as the ones we consider\footnote{In the situations we consider the reduced state of modes $i$ and $j$ is nonsymmetric,
since in general the mixedness $a_i$ and $a_j$ (defined in \eqref{eq.W7}) are not equal.}. 
The logarithmic negativity shares with the  entanglement of formation the drawback of possibly violating monogamy inequalities \cite{Adesso2006}. This is a relatively mild drawback in the context of evaluating bipartite entanglement, but becomes prohibitive in the tripartite context. The Gaussian contangle was introduced in Ref. \cite{Adesso2006} as a quantity which has none of the previous deficiencies. It has been studied in the gravitational context \cite{Adesso2009} and also
in analog gravity \cite{Isoard2021}, but it is not of very practical use as its determination requires numerical minimisation of a complex expression. The PPT measure we use in the present study was introduced in Ref. \cite{Isoard2021} as a measure which mimics many aspects of the Gaussian contangle but is much simpler to evaluate. Its evaluation is 
as simple as the one of the above quoted quantities albeit it 
shares none of their drawbacks. The explicit expression of $\Lambda^{(i|2)}(\omega)$ characterizing the coupling between mode $i=0$ or $1$ and mode $j=2$ is given in Eq. \eqref{eq.entang6} of the main text.

As clear from the above expressions \eqref{eq.apsig2bis}, \eqref{eq.apsig4} and \eqref{eq.apsig5}, the theoretical evaluation of the components of the covariance matrix relies on the computation of averages of two creation or annihilation operators of the outgoing modes. The relevant expressions are determined from Eq. \eqref{eq.excit3} and read ($i=0$ or 1):
\begin{equation}\label{eq.apsig8}
\begin{split}
    \langle \hat{c}^\dagger_i \hat{c}_i\rangle = &
    |S_{i0}|^2 \bar{n}_0 + |S_{i1}|^2 \bar{n}_1 + |S_{i2}|^2 (1+\bar{n}_2),         \\
\langle \hat{c}_2^\dagger \hat{c}_2\rangle=& |S_{20}|^2 \bar{n}_0 
+ |S_{21}|^2 \bar{n}_1 + |S_{22}|^2 (1+\bar{n}_2)-1,\\
\langle \hat{c}_0 \hat{c}_1^\dagger \rangle=& S_{00}S_{10}^* \bar{n}_0 + S_{01}S_{11}^* \bar{n}_1 + S_{02}S_{12}^* (1+\bar{n}_2),\\
\langle \hat{c}_i \hat{c}_2\rangle =& S_{i0}S_{20}^* \bar{n}_0 
+ S_{i1}S_{21}^* \bar{n}_1 + S_{i2}S_{22}^* (1+\bar{n}_2),
        \\\end{split}
\end{equation}
where the $\bar{n}_j$'s ($j=0$, 1 or 2) are the occupation numbers of the incoming modes, see Eq. \eqref{eq.entang5}. 
The coefficients of the $S$ matrix appearing in the above formulae can be determined numerically as explained in Ref. \cite{Larre2012}. 
Expressions \eqref{eq.apsig8} are useful for computing the PPT measure of entanglement \eqref{eq.entang6} and the CHSH
\eqref{eq.GKMR6} and Svetlichny \eqref{eq.TP7} parameters.
As an illustration we now indicate how to compute the PPT measure $\Lambda^{(i|2)}$ at zero temperature. Using here the definition \eqref{eq.GKMR7} as a short hand notation one gets from \eqref{eq.entang6}:
\begin{equation}\label{eq.apsig9bis}
\begin{split}
    \Lambda^{(i|2)}(\omega) = & \sqrt{[|S_{i2}|^2+ \sinh^2(r_2)]^2 + 4 |S_{i2}|^2}\\ & -|S_{i2}|^2-\sinh^2(r_2).
    \end{split} 
\end{equation}
It was shown in Ref. \cite{Larre2012} that the ratio $|S_{i2}|^2/\sinh^2(r_2)$ tends to a constant when $\omega\to 0$. Let's denote $\Gamma_i$ the value of this constant ($i=0$ or 1) \footnote{$\Gamma_0$ and $\Gamma_1$ are the low energy limits of the transmission and reflection coefficients ($\gamma_0$ and $\gamma_1$, respectively) of the beam splitter involved in the effective optical model depicted in Fig. \ref{fig.optical.model}, see Appendix \ref{app:analogue}. In the notations of Appendix \ref{app:analogue}, $\Gamma_0=\lim_{\omega\to 0}\cos^2\theta$.}. A simple expansion of \eqref{eq.apsig9bis} shows that
\begin{equation}\label{eq.apsig98}
    \lim_{\omega\to 0}\Lambda^{(i|2)} =
    \frac{2 \Gamma_i}{1+\Gamma_i}.
\end{equation}
In the waterfall configuration \cite{Larre2012}
\begin{equation}\label{eq.Gamma0}
\Gamma_0=\frac{4 m_u}{(1+m_u)^2},
\end{equation}
and from relation \eqref{eq.excit4} it follows that $\Gamma_0+\Gamma_1=1$. Hence, for the waterfall configuration
\begin{subequations}\label{eq.apsig99}
\begin{align}
&    \lim_{\omega\to 0}\Lambda^{(0|2)} =
    \frac{8 m_u}{1+6m_u+m_u^2}, \label{eq.apsig99a}   \\
&    \lim_{\omega\to 0}\Lambda^{(1|2)} =
    \frac{(1- m_u)^2}{1+m_u^2}. \label{eq.apsig99b}
\end{align}
\end{subequations}
The maximum value of $\Lambda^{(0|2)}$ is always reached at $\omega=0$, thus the $T=0$
numerically determined value ${\rm max}_\omega \Lambda^{(0|2)}$ plotted in Fig. 
\ref{fig3:plus} is identical to \eqref{eq.apsig99a}. The maximum value of the PPT 
measure of entanglement between modes 1 and 2 is reached at $\omega=0$ only for $m_u\lesssim 0.18$: the 
numerically determined value ${\rm max}_\omega \Lambda^{(1|2)}$ plotted in Fig. 
\ref{fig3:plus} is thus identical to \eqref{eq.apsig99b} in this range of values of $m_u$.

\section{An analogue optical system}\label{app:analogue}

As discussed in Ref. \cite{Isoard2021}, the entanglement in the system can be localized by a transformation involving effective modes $\hat{f}_0,\hat{f}_1,\hat{f}_2$ schematically represented in Fig. \ref{fig.optical.model}. These modes are related to the physical outgoing modes by:
\begin{subequations}
 \label{eq.apsig9}
\begin{align}
 \hat{f}_0  &= -\sin \theta\,  \hat{e}_{0} + \cos \theta \, \hat{e}_{1}, \label{eq:expressions_e_theta1}\\
  \hat{f}_1  &= \cos \theta \, \hat{e}_{0} + \sin \theta \, \hat{e}_{1}, \label{eq:expressions_e_theta2}\\
\hat{f}_2&=\hat{e}_{2} \label{eq:expressions_e_theta3},
\end{align}
\end{subequations}
 where
\begin{equation}
\label{eq.apsig10}
\hat{e}_0 = \frac{S_{02}^*}{|S_{02}|}\, \hat{c}_0, \,\,
\hat{e}_1 = \frac{S_{12}^*}{|S_{12}|}\, \hat{c}_1, \,\,
\hat{e}_2 = \frac{S_{22}}{|S_{22}|}\, \hat{c}_2,
\end{equation}
with
\begin{equation}\label{eq.apsig11}
   \cos\theta=\frac{|S_{02}|}{\sinh r_{2}}, \quad
   \sin \theta=\frac{|S_{12}|}{\sinh r_{2}}, 
\end{equation}
and
\begin{equation}\label{eq.GKMR7}
    r_2(\omega)= \operatorname{arsinh}\sqrt{|S_{22}|^2-1}.
\end{equation}
It has been shown in Ref. \cite{Isoard2021} that an incoming vacuum mode of frequency $\omega$ relates to that of the effective $f$ modes by
\begin{equation}\label{eq.apsig11bis}
|0_\omega\rangle^{\rm in} = \exp\left[r_2 ( \hat{f}_1^\dagger \, \hat{f}_2^\dagger - \hat{f}_1 \, \hat{f}_2 )\right]  |0_\omega\rangle^{f},
 \end{equation}
indicating that the $T=0$ state of the system is, as far as the $f_1$ and $f_2$ modes are concerned, a two-mode squeezed vacuum with squeezing parameter $r_2(\omega)$. The $f_1$ and $f_2$ modes are mixed by
a beamsplitter (see Fig. \ref{fig.optical.model}) of transmission and reflection coefficients $\cos^2\theta$ and $\sin^2\theta$, respectively. It has been argued in Ref. \cite{Isoard2021} that 
the long wavelength limit of the transmission coefficient,
$\Gamma_0=\lim_{\omega\to 0} \cos^2\theta$, plays the role of the grey-body factor of the analog black hole.

From definitions \eqref{eq.apsig9} and \eqref{eq.apsig10} and expressions \eqref{eq.apsig8} it is a simple matter
to evaluate the following averages
\begin{equation}\label{eq.apsig12}
\begin{split}
&\langle \hat{f}_0^\dagger \hat{f}_0 \rangle = \sum_{i=0}^1
|S_{12}S_{0i}-S_{02}S_{1i}|^2 \, \bar{n}_i/\sinh^2 r_2,
\\
&    \langle \hat{f}_1^\dagger \hat{f}_1 \rangle =
\cosh^2 r_2 \, \bar{n}_{01} + \sinh^2 r_2 (1+\bar{n}_2), \\
&    \langle \hat{f}_2^\dagger \hat{f}_2 \rangle =
\sinh^2 r_2 \, \bar{n}_{01} + \cosh^2 r_2 (1+\bar{n}_2) -1,
\\
&    \langle \hat{f}_1 \hat{f}_2 \rangle = 
\cosh r_2 \sinh r_2 (\bar{n}_{01}+\bar{n}_2 +1).
\\
\end{split}
\end{equation}
The occupation numbers $\bar{n}_j$ ($j=0$, 1 or 2)  in the above expressions are defined in \eqref{eq.entang5} and use has been made of the shorthand notation
\begin{equation}\label{eq.apsig13}
\bar{n}_{01}\equiv\frac{|S_{20}|^2}{\sinh^2 r_{2}}\,\bar{n}_0 +
\frac{|S_{21}|^2}{\sinh^2 r_{2}}\,\bar{n}_1.
\end{equation}
It is quite informative to quantify the entanglement between the two
effective modes $f_1$ and $f_2$ and the amount by which the
corresponding squeezed state violates Bell inequality. At $T=0$ all
the relevant quantities can be expressed in terms of the squeezing
parameter \eqref{eq.GKMR7} involved in the transformation
\eqref{eq.apsig11bis}. For instance, the analogues for the the modes
$\hat{f}_1$ and $\hat{f}_2$, of the PPT measure of entanglement and of
the CHSH parameter defined for the modes $\hat{c}_i$ and $\hat{c}_2$
by Eqs. \eqref{eq.entang6} and \eqref{eq.GKMR6}, respectively, read
\begin{equation}
    \Lambda^{(f_1|f_2)}=1-\exp(-2 r_2),
\end{equation}
and \cite{Gour2004}    
\begin{equation}\label{eq.GKMR8}    
    B^{(f_1|f_2)}=
    2\sqrt{1+\frac{4}{\pi^2}\arctan^2\left[\sinh(2 r_2)\right]}.
\end{equation}
\begin{figure}
\centering
\includegraphics[width=\linewidth]{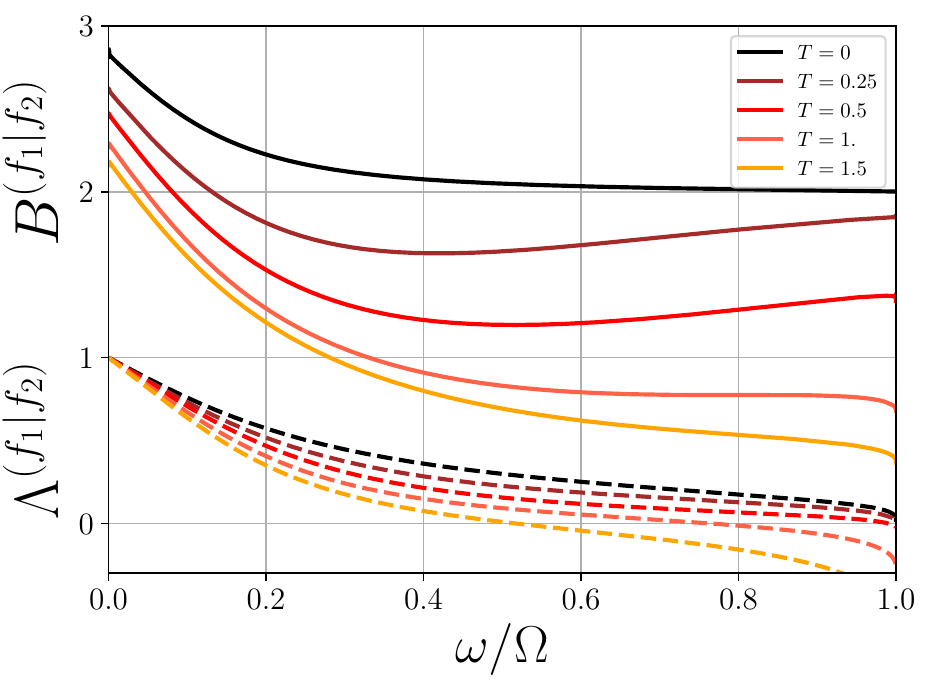}
\caption{$B^{(f_1|f_2)}$ (solid lines) and $\Lambda^{(f_1|f_2)}$
  (dashed lines) plotted as functions of $\omega$ for the two-mode
  squeeze state emulating the the waterfall configuration with
  $m_d=2.9$. The values of the different temperatures are indicated in
  units of $g\,n_u$.}\label{fig3.3}
\end{figure}
The finite temperature values of these quantities can be obtained by
replacing the $\hat{c}$ operators by the $\hat{f}$'s in expressions
\eqref{eq.entang6} and \eqref{GoodCHSHb} and using formulae
\eqref{eq.apsig12}.  They are represented in Fig. \ref{fig3.3} as
functions of $\omega$ for a specific black hole configuration
(waterfall with $m_d=2.9$). As anticipated, it appears that the
violation of Bell inequality is here much more resilient to
temperature than for the true Hawking-partner pair or the
companion-partner pair (compare with Fig. \ref{fig3.1}). This is
explained by the fact that the beam splitter distributes and, so to
say, dilutes the entanglement. This is clear at $T=0$: in this case
the $f_0$ mode is empty [cf. \eqref{eq.apsig12}] and the system is, as
far as the $f$ modes are concerned, in a pure two-mode squeezed vacuum
state. At $\omega=0$ for instance, the system is maximally entangled
(the squeezing parameter $r_2\to\infty$) and both
$\Lambda^{(f_1|f_2)}$ and $B^{(f_1|f_2)}$ reach their upper bounds (1
and $2\sqrt{2}$ respectively).  At variance, if working with the true
outgoing modes described by the $\hat{c}$ (or equivalently the
$\hat{e}$) operators, for studying two-mode entanglement it is
necessary to trace the (occupied) third one. The resulting density
matrix is mixed, and in this case, entanglement and violation of Bell
inequality are two different things, as clear from Fig. \ref{fig3.1}.

Another way to tackle this issue consists in expressing the PPT measure of entanglement \eqref{eq.entang6} between the partner (mode 2) and mode $i$ ($i=0$ or 1) in terms
of the parameters of the equivalent optical system. Denoting as 
$\gamma_0(\omega)=\cos^2\theta$ and 
$\gamma_1(\omega)=\sin^2\theta$ the transmission and reflection coefficient of the beam splitter makes it possible to write, at zero temperature, the PPT measure  \eqref{eq.entang6} under the form
\begin{equation}
\begin{split}
  \Lambda^{(i|2)}(\omega)& = (1+\gamma_i) \sinh r_2\times  \\
& \left\{ \sqrt{\cosh^2 r_2 -
\left(\frac{1-\gamma_i}{1+\gamma_i}\right)^2}
-\sinh r_2 \right\}.
\end{split}
\end{equation}
Algebraic manipulations then show that, because $\gamma_i\le 1$, $\Lambda^{(i|2)}$ is always lower that $\Lambda^{(f_1|f_2)}$. 
This means that the entanglement between modes $i$ and  $2$ is always lower than the entanglement between the modes issued from the parametric down converted represented in Fig. \ref{fig.optical.model}.
The equality is reached when $\gamma_i=1$, and, since $\gamma_0+\gamma_1=1$, in this case the other channel (let's denote it as $\bar{\imath}=1-i$) is not entangled with mode 2: $\Lambda^{(\bar\imath|2)}=0$. So, indeed, the effect of the beam splitter is, so to say, to dispatch the entanglement of the effective squeezed modes $f_1$ and $f_2$ between modes $c_0$, $c_1$ and $c_2$.

After discussing two modes entanglement it is also interesting to briefly consider tripartite entanglement in the effective optical system depicted in Fig. \ref{fig.optical.model}. This system has been designed -- by means of a procedure called entanglement localization \cite{Serafini2005} -- in such a way that 
it concentrates entanglement between the effective squeezed modes $f_1$ and $f_2$.
Therefore, it should be expected that, in the $f$ basis, the quantity $\langle \hat{\mathscr{S}}{}^{(f_0|f_1|f_2)}\rangle$, which is sensible to a genuine tripartite nonlocality, will never reach values above two. Indeed, in this basis, at zero temperature one has
\begin{equation}
 \langle \hat{\Pi}_r^{(f_0)} \otimes  \hat{\Pi}_s^{(f_1)}  \otimes  \hat{\Pi}_t^{(f_2)} \rangle = \langle \hat{\Pi}_r^{(f_0)} \rangle \langle \hat{\Pi}_s^{(f_1)}  \otimes  \hat{\Pi}_t^{(f_2)} \rangle.
\label{theTcoeff_in_f_basis}
\end{equation}
Since
\begin{equation}
\langle \hat{\Pi}_z^{(f_0)} \rangle = 1, \quad 
\langle \hat{\Pi}_x^{(f_0)} \rangle =
\langle \hat{\Pi}_y^{(f_0)} \rangle = 0,
\label{eq.C12}
\end{equation}
it is legitimate to take, for evaluating expression \eqref{eq.TP6} the vector ${\boldsymbol a}_+ = \pm {\boldsymbol e}_z$ (taking $\boldsymbol{a}_- = \pm \boldsymbol{e}_z$ instead doesn't change the result \eqref{B_3modes_fbasis} below). This leads to
\begin{equation}
\begin{split}
\text{max}_{ \theta} \langle \hat{\mathscr{S}}{}^{(f_0|f_1|f_2)}  \rangle = 
 \Big\vert \langle & \boldsymbol{b} \cdot \hat{\boldsymbol\Pi}{}^{(f_1)}  \otimes  \boldsymbol{c}' \cdot \hat{\boldsymbol\Pi}{}^{(f_2)}  \rangle\\
 + & \langle \boldsymbol{b}' \cdot \hat{\boldsymbol\Pi}{}^{(f_1)} \otimes \boldsymbol{c} \cdot \hat{\boldsymbol\Pi}{}^{(f_2)} \rangle \Big\vert .
 \end{split}
\label{}
\end{equation}
Then, noticing that at zero temperature 
$\langle \hat{\Pi}^{(f_1)}_z\otimes  \hat{\Pi}^{(f_2)}_z\rangle =1$
at all energies and that the vector $\boldsymbol{b}$, $\boldsymbol{b}'$, $\boldsymbol{c}$ and $\boldsymbol{c}'$ are not constrained with respect to each other, one can take these vectors to be $+ \boldsymbol{e}_z$, thus obtaining
\begin{equation}
S^{(f_0|f_1|f_2)}(\omega) \underset{T=0} = 2.
\label{B_3modes_fbasis}
\end{equation}

Of course the value of $S^{(f_0|f_1|f_2)}$
decreases
when the temperature increases. This quantity is thus always lower than 2, indicating,  as expected, that there is 
no genuine tripartite nonlocality between the effective $f$ modes.

From  Eqs. \eqref{theTcoeff_in_f_basis}, \eqref{eq.C12} and also from the facts that 
$\langle \hat{\Pi}_s^{(f_1)}  \otimes  \hat{\Pi}_t^{(f_2)} \rangle=0$
if $s\neq t$ and that, at $\omega=0$ and $T=0$,
\begin{equation}
\langle \hat{\Pi}_x^{(f_1)}  \otimes  \hat{\Pi}_x^{(f_2)} \rangle=
-\langle \hat{\Pi}_y^{(f_1)}  \otimes  \hat{\Pi}_y^{(f_2)} \rangle=1,
\end{equation}
it is easily found that, at $\omega=0$, the optimized Mermin parameter \eqref{eq.M1} of the $f$ modes is
\begin{equation}\label{eq.Mermin_f}
M^{(f_0|f_1|f_2)}(0) \underset{T=0} = 2.
\end{equation}
This shows that, at variance with the $c$ modes, the effective $f$ modes do not violate the Mermin-Klyshko inequality and certainly do not exhibit the GHZ paradox. Results \eqref{B_3modes_fbasis} and \eqref{eq.Mermin_f} were expected: since the $f$ modes do not exhibit tripartite entanglement they should violate none of the inequalities \eqref{eq.M3}.

It follows from \eqref{eq.Gamma0} that when $m_u=0$ or 1, at $\omega=0$, $\cos\theta=0$ or 1, respectively. 
In this case Eqs. \eqref{eq.apsig9}, \eqref{eq.apsig10} 
and \eqref{eq.apsig11} indicate that
the $f$ modes are connected to the $c$ modes by LLUBOs:
the tripartite character of the true system thus disappears in these two limiting cases, even when $T=0$. However our results indicate that these 
two limits are singular, since the GHZ character of the system is observed (at $\omega=0$ and $T=0$) for all $m_u\in]0,1[$.

\section{Eigenstates of the pseudo-spin operators}\label{app.pseudo.spin}

In this appendix we list the properties of the eigenstates of the pseudo-spin operators \eqref{eq.GKMR1} which are useful in the main text. We will consider a given one of the three modes ($j=0$, 1 or 2), always the same, and will omit the associated label $(j)$ in order to lighten the notations. In a similar way, we do not write the $\omega$-dependence which is implicit in all this appendix.

From the definition \eqref{eq.GKMR1c} it follows that the matrix elements of $\hat{\Pi}_z$ between two number states are
\begin{equation}\label{eq.ps1}
    \langle n | \hat{\Pi}_z |m \rangle =
\left[(-)^n+(-)^m\right] \int_0^\infty {\rm d}q\, \psi_n(q)\psi_m(q),
\end{equation}
where $\psi_n(q)=\langle n|q\rangle
=
(2^n n! \sqrt{\pi}\,)^{-1/2} \exp(-q^2/2) H_n(q)$ is a normalized Hermite function ($H_n$ a Hermite polynomial). The prefactor in the r.h.s. of 
\eqref{eq.ps1} imposes that $n$ and $m$ have the same parity. Thus $\psi_n$ and $\psi_m$ are both even or odd functions of $q$, which makes it possible to extend by symmetry the integration range in the r.h.s. of \eqref{eq.ps1} to the whole real axis, leading to
\begin{equation}\label{eq.ps2}
    \langle n | \hat{\Pi}_z | m \rangle=(-)^{n} \delta_{n,m},
\end{equation}
and thus
\begin{equation}\label{eq.ps3}
    \hat{\Pi}_z = \sum_{n=0}^\infty \Big(
    |2n\rangle \langle 2n| - |2n+1\rangle \langle 2n+1|
    \Big):
\end{equation}
the eigenstates of $\hat{\Pi}_z$ are the number states, they have eigenvalue $\pm 1$ depending on their parity. In the following we will denote them as
\begin{equation}\label{eq.ps4a}
|z^{+}_{n}\rangle=|2n\rangle, \quad\mbox{and}\quad |z^{-}_{n}\rangle=|2n+1\rangle. 
\end{equation}
Contrarily to usual spins 1/2, the two eigenvalues (here $+1$ and $-1$) are infinitely degenerate. This property is shared by any projection of the pseudo-spin operator; it is in particular true for the operator $\hat{\Pi}_x$. We denote the eigenstates of $\hat{\Pi}_x$ associated to the eigenvalue $\pm 1$  by
$|x^{\pm}_{n}\rangle$, they
can be constructed 
by rotating the
eigenstates $|z^{\pm}_{n}\rangle$ of $\hat{\Pi}_z$ by angles $\pm \pi/2$ around the $y$ axis:
\begin{equation}\label{eq.ps4}
|x^{\pm}_{n}\rangle = \hat{\cal R}_y\left(\pm \frac{\pi}{2}\right) 
   |z^{\pm}_{n}\rangle =
   \frac{1}{\sqrt{2}} \left(\mathds{1} \mp \hat{\Pi}_y\right)
   |z^{\pm}_{n}\rangle,
\end{equation}
where $\hat{\cal R}_y(\theta)=\exp(-\frac{i}{2} \theta \, \hat{\Pi}_y)$ is the operator of rotation of angle $\theta$ around the $y$ axis. The fact that
$|x^{\pm}_{n}\rangle$ is an eigenstate of $\hat{\Pi}_x$ with eigenvalue $\pm 1$ is easily checked by direct application of $\hat{\Pi}_x$ to the left of expression \eqref{eq.ps4} and use of the relations
\begin{equation}\label{eq.ps6}
    \hat{\Pi}_r \hat{\Pi}_s = i \, \varepsilon_{rst}\, \hat{\Pi}_t,
\end{equation}
where $\varepsilon_{rst}$ is the totally antisymmetric Levi-Civita symbol and $(r,s,t)=x$, $y$ or $z$. By the same token it is also easily proven that
\begin{equation}\label{eq.ps7}
    \hat{\Pi}_y |x^{\pm}_{n}\rangle =
    \mp i |x^{\mp}_{n}\rangle, \quad\mbox{and}\quad
\hat{\Pi}_z |x^{\pm}_{n}\rangle =
    |x^{\mp}_{n}\rangle. 
\end{equation}

\section{Computing expectation values of observables}\label{app.averages}

The use of the technique of Wigner transform (see, e.g., Refs. \cite{Case2008,Scheich2001}) is particularly well suited for
determining the different contributions of the pseudo-spin operator to the CHSH measures \eqref{eq.GKMR6} and \eqref{eq.TP7}. The reason is twofold: first, we consider a Gaussian state and for computing the required averages we thus just need to evaluate Gaussian integrals weighted by the Wigner transforms of the spin operators and secondly, the Wigner transforms of the pseudo-spins \eqref{eq.GKMR1} are simple enough that the relevant integrals can be evaluated analytically.

We will consider $n=3$ (and also $n=2$) modes Gaussian states and the computation of the average of an operator $\hat{A}$ is performed in an abstract dimensional phase space according to 
\begin{equation}\label{eq.W1}
    \langle \hat{A}\rangle = \int {\rm d}^n\boldsymbol{q}{\rm d}^n\boldsymbol{p}\,
    W_{\hat{\rho}}(\boldsymbol{q},\boldsymbol{p}) \,
    W_{\!\!\hat{A}}(\boldsymbol{q},\boldsymbol{p}) .
\end{equation}
In this expression $W_{\hat{\rho}}$ is the Wigner transform of the density matrix. In the Gaussian case we consider it can be computed from the knowledge of the covariance matrix \cite{weedbrook2012}:
\begin{equation}\label{eq.W4}
W_{\hat{\rho}}(\boldsymbol{q},\boldsymbol{p}) = \frac{1}{\pi^n \sqrt{{\rm det} \sigma}} \exp\{-\tfrac12 \boldsymbol{\xi}^{\rm \sss T } \sigma^{-1} \boldsymbol{\xi} \},
\end{equation}
where $\sigma$ is the total (or reduced, as appropriate) covariance
matrix defined in Eq. \eqref{eq.entang4}. In this expression, $n=3$
and
$\boldsymbol{\xi} = \sqrt2 \, (q_0,p_0,q_{1},p_{1},q_{2},p_2)^{\rm
  \sss T }$ in the three-mode case. In the reduced two-mode case,
$n=2$ and one should remove from the expression of $\boldsymbol{\xi}$
the entries corresponding to the subscript of the traced mode.  The
Wigner transform \eqref{eq.W4} of a Gaussian state is non-negative,
and it was originally considered impossible to violate Bell’s
inequality under such conditions \cite{Bell1986}. This was latter
proven incorrect \cite{Johansen1997,Banaszek1998}. In particular,
Revzen {\it et al.} \cite{Revzen2005} proved that observables can be
associated to the violation of Bell inequality over a Gaussian state
provided their Wigner transform takes values different from the
eigenvalues of the associated quantal operator. We will see that the
pseudo-spins \eqref{eq.GKMR1} we consider belong to this class of
observables, denoted as ``improper'' in Ref. \cite{Revzen2005}.

The other term involved in the integration \eqref{eq.W1} is the Wigner
transform $W_{\!\!\hat{A}}$ of the operator $\hat{A}$. It is a
function defined in phase space by means of an integration over
``representation space''
\begin{equation}\label{eq.W2}
W_{\!\!\hat{A}}(\boldsymbol{q},\boldsymbol{p}) =
\int {\rm d}^n\boldsymbol{z} \exp\{i \boldsymbol{p}\cdot\boldsymbol{z}\}
\langle \boldsymbol{q}-\tfrac12 \boldsymbol{z} | \hat{A} |
\boldsymbol{q}+\tfrac12 \boldsymbol{z}\rangle.
\end{equation}  
In this expression $\boldsymbol{q}$ (as well as $\boldsymbol{z}$ and $\boldsymbol{p}$) is a vector in an abstract $n=3$ dimensional space (with basis $\boldsymbol{e}_0,\boldsymbol{e}_1,\boldsymbol{e}_2$). 
The kets involved in \eqref{eq.W2} are of the type $|\boldsymbol{Q}\rangle=|
Q_0\rangle_0 \otimes |Q_1\rangle_1 \otimes |Q_2\rangle_2$ where $\boldsymbol{Q}
=Q_0 \boldsymbol{e}_0 + Q_1 \boldsymbol{e}_1 + Q_2 \boldsymbol{e}_2$ and
$|Q\rangle_j$ is the eigenstate of operator $\hat{q}_j$ associated to the eigenvalue $Q$ ($j=0$, 1 or 2). In the case of a reduced two modes Gaussian state 
$n=2$ and the vector $\boldsymbol{Q}$ is two-dimensional: the component associated to the traced mode disappears, as it also does in  $|\boldsymbol{Q}\rangle$.

For evaluating expectation values such as those appearing in
Eqs. \eqref{eq.GKMR6} and \eqref{eq.TP7} we need to compute the Wigner
transforms of the components of the pseudo-spin operators. These are
to be evaluated in a 2 dimensional phase space, since operator
$\hat{\boldsymbol{\Pi}}{}^{(j)}$ concerns a single mode (mode
$j$). The result is independent of $j$ and reads:
\begin{equation}\label{eq.W3}
\begin{split}
W_{\hat{\Pi}_x}(q,p)= & \mbox{sgn}\, (q), \\
W_{\hat{\Pi}_y}(q,p)= & i \delta(q)
\int_{-\infty}^\infty\!\!{\rm d}x \, \mbox{sgn}\,(x) \exp\{-2 i p x\} \\
=& \delta(q) {\cal P}(1/q), \\
W_{\hat{\Pi}_z}(q,p)= & \pi \delta(q)\delta(p) ,
\end{split}
\end{equation}
where ${\cal P}$ denotes the principal value. It is shown in the main
text that the eigenvalues of the projections along a given axis of the
pseudo-spin operator $\hat{\boldsymbol{\Pi}}$ are $\pm 1$. The above
expressions of the Wigner transforms thus demonstrate that, contrarily
to $\hat{\Pi}_x$, operators $\hat{\Pi}_y$ and $\hat{\Pi}_z$ are
``improper'' in the sense of Ref. \cite{Revzen2005}: they may be
involved in violation of Bell inequality even for a state with a
non-negative Wigner transform, such as the Gaussian state we consider.

For evaluating the expectation values appearing in \eqref{eq.GKMR6} we
need to compute integrals such as
\begin{equation}\label{eq.W5}
\begin{split}
\langle \hat{\Pi}^{(i)}_z \otimes \hat{\Pi}^{(2)}_z\rangle
=\int & \!{\rm d}^2\boldsymbol{q}{\rm d}^2\boldsymbol{p}\,
W_{\hat{\Pi}_z}(q_i,p_i)\times \\
& W_{\hat{\Pi}_z}(q_2,p_2)
W_{\hat{\rho}}(\boldsymbol{q},\boldsymbol{p})  ,  
\end{split}
\end{equation}
where $\boldsymbol{q}=(q_i,q_2)$ and $\boldsymbol{p}=(p_i,p_2)$ ($i=0$ or 1). 
The explicit calculation yields
\begin{equation}\label{eq.W6}
\langle \hat{\Pi}^{(i)}_z \otimes \hat{\Pi}^{(2)}_z\rangle
=\frac{1}{a_i a_2 -  4|\langle \hat{c}_i\hat{c}_2\rangle|^2} ,
\end{equation}
where 
\begin{equation}\label{eq.W7}
    a_j=2 \langle \hat{c}^\dagger_j\hat{c}_j\rangle + 1
\end{equation}
is known as the local mixedness of mode $j$ ($j=0,1$ or 2). The other terms involved in the determination of the CHSH parameter \eqref{eq.GKMR6} can be evaluated by the same technique. The computation is similar to the one figuring in the Appendix of Ref. \cite{Martin2017}. One finds that 
all expectation values for which the index $z$ appears a single time cancel at all temperatures. The values of the other non-zero averages are  ($i=0$ or 1):
\begin{equation}\label{eq.W8_1}
\begin{split}
\langle \hat{\Pi}^{(i)}_x \otimes \hat{\Pi}^{(2)}_x\rangle = & \frac{2}{\pi}
\arctan \frac{2\, \mbox{Re}\, \langle\hat{c}_i\hat{c}_2\rangle}{\sqrt{a_i a_2 - 
    4(\mbox{Re}\, \langle \hat{c}_i\hat{c}_2\rangle)^2}}, \\
\langle \hat{\Pi}^{(i)}_y \otimes \hat{\Pi}^{(2)}_y\rangle = &
\frac{-1}{A_{i2}}
\langle \hat{\Pi}^{(i)}_x \otimes \hat{\Pi}^{(2)}_x\rangle, \\
\langle \hat{\Pi}^{(i)}_x \otimes \hat{\Pi}^{(2)}_y\rangle = &
\frac{2}{\pi\,a_2}\operatorname{arsinh}
\frac{2\, \mbox{Im}\, \langle\hat{c}_i\hat{c}_2\rangle}{\sqrt{A_{i2}}},\\
\langle \hat{\Pi}^{(i)}_y \otimes \hat{\Pi}^{(2)}_x\rangle = &
\frac{2}{\pi\,a_i}\operatorname{arsinh}
\frac{2\, \mbox{Im}\, \langle\hat{c}_i\hat{c}_2\rangle}{\sqrt{A_{i2}}},\\
\end{split}
\end{equation}
where $A_{i2}$ is defined in Eq. \eqref{eq.W10} below, and we recall that the expression of the quantity $\langle\hat{c}_i\hat{c}_2\rangle$ is given in \eqref{eq.apsig8}.

For studying the Svetlichny observable it is necessary to evaluate averages involving the Cartesian coordinates of three pseudo-spins, of the type
\begin{equation}\label{eq.W9}
    {\cal T}_{rst}(\omega) \equiv \left\langle  \hat{\Pi}^{(0)}_r \otimes \hat{\Pi}^{(1)}_s
    \otimes \hat{\Pi}^{(2)}_t\right\rangle,
\end{equation}
where $r$, $s$ and $t\in\{x,y,z\}$. These quantities are zero if $z$ is not one of the indices or appears exactly twice. 
In order to write down tractable expressions in the other cases, it is convenient to introduce new compact notations:
\begin{equation}\label{eq.W10}
\begin{split}
A_{01}=& a_0a_1- 4\,|\langle \hat{c}_0\hat{c}_1^\dagger\rangle|^2,\\
A_{i2}\underset{i<2}{=} & a_ia_2- 4\,
|\langle \hat{c}_i\hat{c}_2\rangle|^2,
\end{split}
\end{equation}
\begin{equation}\label{eq.W11}
\begin{split}
&Z_0=-2 a_0 \langle \hat{c}_1\hat{c}_2\rangle^*
+ 4 \langle \hat{c}_0\hat{c}_1^\dagger\rangle 
\langle \hat{c}_0\hat{c}_2\rangle^*
,\\
&Z_1=-2 a_1 \langle \hat{c}_0\hat{c}_2\rangle^*
+ 4 \langle \hat{c}_0\hat{c}_1^\dagger\rangle^* 
\langle \hat{c}_1\hat{c}_2\rangle^*
,\\
&Z_2=-2 a_2 \langle \hat{c}_0\hat{c}_1^\dagger\rangle^*
+ 4 \langle \hat{c}_1\hat{c}_2\rangle 
\langle \hat{c}_0\hat{c}_2\rangle^*
,
\end{split}
\end{equation}
and
\begin{equation}\label{eq.W12}
\begin{split}
    \delta= & a_0 a_1 a_2 +16 \, \mbox{Re}\left\{
   \langle \hat{c}_0\hat{c}^\dagger_1\rangle
   \langle \hat{c}_1 \hat{c}_2\rangle
   \langle \hat{c}_0 \hat{c}_2\rangle^*
    \right\}\\
&    -4 a_0 |\langle \hat{c}_1 \hat{c}_2\rangle|^2
    -4 a_1 |\langle \hat{c}_0 \hat{c}_2\rangle|^2
    - 4 a_2 |\langle \hat{c}_0\hat{c}^\dagger_1\rangle|^2.
\end{split} 
\end{equation}
$\delta$ is the square of the determinant of the $6\times 6$ covariance matrix \eqref{eq.apsig1}. At $T=0$ the system is in a pure state and $\delta=1$ for all values of $\omega$ \cite{note.delta1}, whereas $\delta={\cal O}(1/\omega)$ at finite temperature, as can be shown on the basis of the low energy expansion of the matrix elements of the $S$-matrix given in Ref. \cite{Larre2012}.

The explicit theoretical evaluation of the quantities \eqref{eq.W7}, \eqref{eq.W10}, \eqref{eq.W11} and \eqref{eq.W12} is easily done from Eqs. \eqref{eq.apsig8}.
A long computation shows that the averages \eqref{eq.W9} which are non zero can be expressed 
in terms of these quantities:
\begin{equation}\label{eq.W13}
\begin{split}
&    {\cal T}_{zxx}=-\frac{2}{\pi a_0}
    \arctan\frac{\mbox{Re}\, Z_0}{\sqrt{A_{01}A_{02}-(\mbox{Re}\, Z_0)^2}}
,\\
&    {\cal T}_{xzx}=-\frac{2}{\pi a_1}
    \arctan\frac{\mbox{Re}\, Z_1}{\sqrt{A_{01}A_{12}-(\mbox{Re}\, Z_1)^2}}
,\\
&    {\cal T}_{xxz}=-\frac{2}{\pi a_2}
    \arctan\frac{\mbox{Re}\, Z_2}{\sqrt{A_{02}A_{12}-(\mbox{Re}\, Z_2)^2}}
,
\end{split}
\end{equation}
\begin{equation}\label{eq.W14}
\begin{split}
& {\cal T}_{zyy}=-\frac{a_0}{\delta} {\cal T}_{zxx},\quad
{\cal T}_{yzy}=-\frac{a_1}{\delta} {\cal T}_{xzx},\\
& {\cal T}_{yyz}=+\frac{a_2}{\delta} {\cal T}_{xxz},
\end{split}
\end{equation}
\begin{equation}\label{eq.W14b}
    \begin{split}
 A_{02}{\cal T}_{zxy}=A_{01}{\cal T}_{zyx}=&
    \operatorname{arsinh}\left(\frac{\mbox{Im}\,Z_0}{\sqrt{a_0\delta}}\right),\\
 A_{01}{\cal T}_{yzx}=A_{12}{\cal T}_{xzy}=&
    \operatorname{arsinh}\left(\frac{\mbox{Im}\,Z_1}{\sqrt{a_1\delta}}\right),\\
 -A_{12}{\cal T}_{xyz}=A_{02}{\cal T}_{yxz}=&
    \operatorname{arsinh}\left(\frac{\mbox{Im}\,Z_2}{\sqrt{a_2\delta}}\right),
    \end{split}
\end{equation}
and
\begin{equation}\label{eq.W15}
    {\cal T}_{zzz}=\frac{1}{\delta}.
\end{equation}
The low energy behavior of the quantities \eqref{eq.W13} is dictated by the one of the local mixednesses $a_0$, $a_1$ and $a_2$ which diverge as $1/\omega$ for all temperature. As a result
\begin{equation}\label{eq.lim1}
    \lim_{\omega\to 0} {\cal T}_{zxx}=\lim_{\omega\to 0} {\cal T}_{xzx}
    =\lim_{\omega\to 0} {\cal T}_{xxz}
    \underset{\forall T}{=}0.
\end{equation}
On the other hand, the behavior of the quantities \eqref{eq.W14} depends on the behavior of the (square root $\delta$ of the) 
determinant of the covariance matrix which is unity at $T=0$ \cite{note.delta1}. In this case 
\begin{equation}\label{eq.lim2z}
{\cal T}_{zzz}(\omega)\underset{T = 0}{=}1,
\end{equation}
and
\begin{equation}\label{eq.lim2}
    \lim_{\omega\to 0} {\cal T}_{zyy}=\lim_{\omega\to 0} {\cal T}_{yzy}
    =\lim_{\omega\to 0} {\cal T}_{yyz}\underset{T = 0}{=} 1.
\end{equation}
The reason for this behavior is that, at $T=0$, $\delta=1$ whereas
the arguments of all the $\arctan$ terms in \eqref{eq.W13} diverge as $\omega^{-1/2}$, as shown by detailed inspection based on
Eqs. \eqref{eq.W10}, \eqref{eq.W11}, \eqref{eq.apsig8} and the asymptotic expression of the coefficients of the $S$ matrix given in \cite{Larre2012}.

Alternatively, at finite temperature $\delta$ diverges at low energy (as $1/\omega$) and the limits \eqref{eq.lim2} all cancel
\begin{equation}\label{eq.lim3}
    \lim_{\omega\to 0} {\cal T}_{zyy}=\lim_{\omega\to 0} {\cal T}_{yzy}
    =\lim_{\omega\to 0} {\cal T}_{yyz}
    \underset{T \ne 0}{=} 0,
\end{equation}
as also does $\lim_{\omega \to 0} {\cal T}_{zzz}\underset{T \ne 0}{=}0$.

\section{Analytic maximization of $\langle \hat{\mathscr{B}}{}^{(i|2)} \rangle$} \label{app.maxi2}

In this appendix we present the maximisation of the expectation value of the CHSH operator \eqref{eq.GKMR3}. We only state the results useful for the main text but do not detail the procedure because it is well known, see e.g., Refs. \cite{Popescu1992,Horodecki1995}.
In particular, it can be shown that the CHSH parameter $B^{(i|2)}(\omega)$ defined in  \eqref{eq.GKMR6} reads
\begin{equation}\label{eq.am2_1}
    B^{(i|2)}=2\sqrt{\lambda_1+\lambda_2} \, ,
\end{equation}
where $\lambda_1$ and $\lambda_2$ are the two largest eigenvalues of 
matrix $^{\rm \sss T }{\cal T}\, {\cal T}$ where ${\cal T}(\omega)$ is the $3\times 3$ matrix with entries 
\begin{equation}
{\cal T}_{rs}
=\left\langle \hat{\Pi}^{(i)}_r \otimes \hat{\Pi}^{(2)}_s\right\rangle ,
\end{equation}
with $(r,s)\in\{x,y,z\}^2$. The results presented in Appendix \ref{app.averages} show that in all the cases we consider the matrix
${\cal T}(\omega)$ is block diagonal:
\begin{equation}\label{eq.am2_2}
    {\cal T}=\begin{pmatrix}
        {\cal T}_{xx} & {\cal T}_{xy} & 0 \\
        {\cal T}_{yx} & {\cal T}_{yy} & 0 \\
        0 & 0 & {\cal T}_{zz} \\
    \end{pmatrix},
\end{equation}
and thus
\begin{equation}\label{eq.am2_2bis}
^{\rm \sss T }{\cal T}\,  {\cal T}=\begin{pmatrix}
        A & C & 0 \\
        C & B & 0 \\
        0 & 0 & {\cal T}^2_{zz} \\
    \end{pmatrix},
\end{equation}
with
\begin{equation}\label{eq.am2_3}
\begin{split}
A& = {\cal T}_{xx}^2
+ {\cal T}_{yx}^2,\quad
B ={\cal T}_{yy}^2
+ {\cal T}_{xy}^2,\\ 
C&= {\cal T}_{xx} {\cal T}_{xy}+{\cal T}_{yy} {\cal T}_{yx}.
\end{split}    
\end{equation}
The largest eigenvalues of matrix $^{\rm \sss T }{\cal T}\, {\cal T}$ are
${\cal T}_{zz}^2$ and
\begin{equation}\label{eq.am2_4}
    \frac12 \left( A+B + \sqrt{(A-B)^2+4 C^2} \right).
\end{equation}
They can be computed from Expressions \eqref{eq.W6}, \eqref{eq.W8_1} and \eqref{eq.am2_3}. Then Eq. \eqref{eq.am2_1} determines the value of the CHSH parameter \eqref{eq.GKMR6}.

One may also chose a different strategy for attempting 
to maximise the expectation value of the operator $\mathscr{B}^{(i|2)}$. Instead of choosing to work in the basis of the $c$-
modes, one may perform a LLUBO for attempting to simplify the form of the covariance matrix.
As already stated in Appendix \ref{app.sigma}, in the bipartite case, the $4\times 4$ covariance
matrix \eqref{eq.apsig7} associated with the reduced two-mode state $(i|2)$ can be brought by LLUBOs to a standard form where the matrix $\varepsilon_{i2}$ are diagonal \cite{Duan2000}. Working in this basis does not alter the entanglement properties of the system (they remain unaffected compared to that of the $c$-modes), but makes the computations easier and may improve the signal of nonlocality. In this basis the result \eqref{eq.W6} is not affected and expressions
\eqref{eq.W8_1} modify to
\begin{equation}\label{eq.W8_2}
\begin{split}
{\cal T}_{xx}
= \;  & \frac{2}{\pi}
\arctan \frac{2\, | \langle\hat{c}_i\hat{c}_2\rangle |}{\sqrt{A_{i2}}}= -A_{i2} {\cal T}_{yy}, \\
{\cal T}_{xy} = \; &
{\cal T}_{yx} = 
0,\\
\end{split}
\end{equation}
where $A_{i2}$ is defined in Eq. \eqref{eq.W10}. The matrix ${\cal T}$ is thus diagonal and expression \eqref{eq.am2_4} is equal to ${\cal T}_{xx}^2$. Equation \eqref{eq.am2_1} then reads
\begin{subequations}\label{GoodCHSH}
\begin{align}
B^{(i|2)} =  & 2\, \sqrt{{\cal T}_{xx}^2+{\cal T}_{zz}^2}
\label{GoodCHSHa}
\\
=  &
2 \, \sqrt{\frac{4}{\pi^2}
\arctan^2\left(\frac{2|\langle \hat{c}_i\hat{c}_2\rangle|}{\sqrt{A_{i2}}}\right)
+ \frac{1}{A_{i2}^2}}\, .\label{GoodCHSHb}
\end{align}
\end{subequations}
We note here that the difference between the result  \eqref{eq.am2_1} evaluated in the $c$-mode basis and expression \eqref{GoodCHSHb} is always small. The reason is that, in the $c$-mode basis, the off diagonal  entries of the upper left blocks of matrix \eqref{eq.am2_2} and \eqref{eq.am2_2bis} are always small compared to the diagonal ones, because at all temperature 
$|\mbox{Im}\, \langle\hat{c}_i\hat{c}_2\rangle|\ll 
|\mbox{Re}\, \langle\hat{c}_i\hat{c}_2\rangle|$.
 However, our numerical checks always demonstrate a small increase of the Bell parameter \eqref{GoodCHSHb} compared to the one evaluated using \eqref{eq.am2_1} in the $c$-mode basis. We thus present our numerical results in Figs.  \ref{fig3.1}, \ref{fig3:plus}, \ref{fig3.pluT}, 
\ref{fig3.3}, \ref{fig.H1} and \ref{fig.H2} using formula \eqref{GoodCHSHb}. To summarize, the result \eqref{GoodCHSHb} should be considered as an optimized value of the witness of nonlocality $B^{(i|2)}$.
We note here that we do not have a general proof of the better efficiency of the method which consists in using the basis in which the covariance matrix is in its standard form, but we believe that a general mathematical result of this type would be quite useful.

\section{Numerical maximization of $\langle \hat{\mathscr{S}}{}^{(0|1|2)} \rangle$} \label{app.maxi3}

As explained in Sec.~\ref{sec.Bell3}, the maximal value of the three-mode Bell operator $ \langle \hat{\mathscr{S}}{}^{(0|1|2)} \rangle$ can be found by optimizing the orientation of the unit vectors 
$\boldsymbol{a}$, 
$\boldsymbol{a}'$, $\boldsymbol{b}$, $\boldsymbol{b}'$, $\boldsymbol{c}$ and  $\boldsymbol{c}'$. 
However, solving this optimization problem analytically proves challenging due to the need to maximize a function depending on 12 real parameters. Indeed, the orientation of each of the six previous normalized vectors in the 3-dimensional physical space corresponds to two degrees of freedom, leading in total to twelve parameters. 

Consequently, we resort to a numerical method to evaluate the maximal violation of Bell inequalities and determine the corresponding optimal orientations for the vectors $\boldsymbol{a}$, $\boldsymbol{a}'$, $\boldsymbol{b}$, $\boldsymbol{b}'$, $\boldsymbol{c}$, and $\boldsymbol{c}'$. More explicitly, we use a genetic algorithm which has proved very efficient for optimizing 
a function over a large parameter space \cite{DeJong1997}. This algorithm is based on natural selection: the code starts with a 
random set of solutions (in our case each of them consists of 12 
parameters) which form all together what we call a population. We then compute, for each set of vectors 
($\boldsymbol{a}, \boldsymbol{a}', \boldsymbol{b}, \boldsymbol{b}', \boldsymbol{c}, \boldsymbol{c}'$), the expectation value 
$ \langle \hat{\mathscr{S}}{}^{(0|1|2)} \rangle$ by means of the technique exposed in Appendix \ref{app.averages}. For instance the contribution of the term
 $\langle  \boldsymbol{a} \cdot \hat{\boldsymbol{\Pi}}{}^{(0)} \otimes 
\boldsymbol{b} \cdot \hat{\boldsymbol{\Pi}}{}^{(1)} \otimes 
\boldsymbol{c}\cdot \hat{\boldsymbol\Pi}{}^{(2)} \rangle$ to 
$ \langle \hat{\mathscr{B}}{}^{(0|1|2)} \rangle$
 can be evaluated from the knowledge of the terms ${\cal T}_{rst}$ defined in Eq. \eqref{eq.W9}:
\begin{equation}\label{eq.NUM}
    \langle  \boldsymbol{a} \cdot \hat{\boldsymbol{\Pi}}{}^{(0)} \otimes 
\boldsymbol{b} \cdot \hat{\boldsymbol{\Pi}}{}^{(1)} \otimes 
\boldsymbol{c}\cdot \hat{\boldsymbol\Pi}{}^{(2)} \rangle=\sum_{r,s,t} a_{r} b_s c_t {\cal T}_{rst},
\end{equation}
where the sum runs over the indices $(r,s,t)\in\{x,y,z\}^3$. At variance with the bipartite case, at finite temperature it not not possible to find a LLUBO enabling to cast the 
$6\times 6$ covariance matrix \eqref{eq.apsig1} under a standard form where all the 
$\varepsilon_{ij}$ matrices are diagonal (see the discussion in Appendix \ref{app.sigma}). The expectation values are thus computed in the natural basis of the $c$-modes where the values of the ${\cal T}_{rst}$ coefficients are given by  Eqs. \eqref{eq.W13}, \eqref{eq.W14}, \eqref{eq.W14b} 
and \eqref{eq.W15}.

At each step of the algorithm the code computes the expectation value 
$ \langle \hat{\mathscr{S}}{}^{(0|1|2)} \rangle$ for a set of vectors ($\boldsymbol{a}, \boldsymbol{a}', \boldsymbol{b}, \boldsymbol{b}', \boldsymbol{c}, \boldsymbol{c}'$) and ranks the members of the population by computing a \textit{fitness scaling function}, a kind of selection rule: only the members of the population with the lowest fitness value will be retained -- they are called the parents -- and used to generate new sets of solutions, called the children. Then, at the next step, the selection rules are applied to the children, some of them become parents in turn and engender a new generation. The algorithm stops when all children look like their parents, or, in other words, when
\begin{equation}
    |\boldsymbol{v}_{\ell+1} - \boldsymbol{v}_{\ell}| < \delta,
\end{equation}
where $\boldsymbol{v}_{\ell} = (\boldsymbol{a}, 
\boldsymbol{a}', \boldsymbol{b}, \boldsymbol{b}',
\boldsymbol{c},\boldsymbol{c}')$ is the set of solutions at step $\ell$ of the algorithm, and $\delta$ is the chosen convergence precision fixed prior the beginning of the selection process.

In our case, the fitness scaling function is simply the opposite of the average $ \langle \hat{\mathscr{S}}{}^{(0|1|2)} \rangle$ of the three-mode Bell operator. Trying to obtain the lowest fitness score is thus equivalent to maximize the Bell operator. For a given set of vectors $\boldsymbol{v}_{\ell}$ at step $\ell$, the next generation is computed as follows: $ \boldsymbol{v}_{\ell+1} = \boldsymbol{v}_{\ell} + \boldsymbol{w}_{\ell}$, where $\boldsymbol{w}_{\ell}$ is a random weight which controls the mutations between the parents $\boldsymbol{v}_{\ell}$ and the children $\boldsymbol{v}_{\ell+1}$, and which tends to decrease when the code starts to converge. For a detailed presentation of the algorithm we refer to Ref. \cite{Beyer2002}.
Note finally that the procedure just presented is also used for determining the optimized Mermin parameter \eqref{eq.M1}.

\section{Tripartite Cirel'son bound and analytic maximization of $\langle\hat{\mathscr{S}}{}^{(0|1|2)}(\omega=0)\rangle$} \label{app.Maximization}

In this section we first study the upper bound of the quantity $S^{(0|1|2)}(\omega)$ defined in \eqref{eq.TP7} and then study the possibility of maximisation of the average $\langle\hat{\mathscr{S}}{}^{(0|1|2)}\rangle$ of the operator \eqref{eq.TP0} by an appropriate choice of the measurement directions $\boldsymbol{a}$, $\boldsymbol{a}'$, $\boldsymbol{b}$, $\boldsymbol{b}'$, $\boldsymbol{c}$ and $\boldsymbol{c}'$.

Denoting as $\hat{A}=\boldsymbol{a}\cdot\hat{\boldsymbol{\Pi}}{}^{(0)}$, 
$\hat{A}'=\boldsymbol{a}'
\cdot\hat{\boldsymbol{\Pi}}{}^{(0)}$,
$\hat{B}=\boldsymbol{b}\cdot\hat{\boldsymbol{\Pi}}{}^{(1)}$, etc. makes it possible to write the square of the tripartite Svetlichny operator \eqref{eq.TP0} as
\begin{equation}\label{eq.C3.1}
\begin{split}
    4 \left(\hat{\mathscr{S}}{}^{(0|1|2)}\right)^2=& 8 + 
    \{\hat{A},\hat{A}'\}\otimes \{\hat{B},\hat{B}'\}\otimes \{\hat{C},\hat{C}'\}\\
    & - 2 [\hat{A},\hat{A}']\otimes[\hat{B},\hat{B}']\otimes \mathds{1}^{(2)}\\
& - 2 \mathds{1}^{(0)}\otimes [\hat{B},\hat{B}']\otimes [\hat{C},\hat{C}']\\
& - 2 [\hat{A},\hat{A}'].\otimes \mathds{1}^{(1)}\otimes [\hat{C},\hat{C}'],
\end{split}
\end{equation}
where $[\cdot,\cdot]$ and $\{\cdot,\cdot\}$ denote the commutator ans the anticommutator, respectively.
From the SU(2) algebra of the pseudo-spins it is easily proven that
\begin{subequations}\label{eq.C3.2}
\begin{align}
  [\hat{A},\hat{A}']& =2 i (\boldsymbol{a}\times\boldsymbol{a}')\cdot \hat{\boldsymbol{\Pi}}{}^{(0)}, \label{eq.C3.2a} \\
 \{\hat{A},\hat{A}'\}& =2 (\boldsymbol{a}\cdot\boldsymbol{a}') \mathds{1}^{(0)},\label{eq.C3.2b}
 \end{align}
\end{subequations}
with similar formulae for the quantities $\hat{B}$, $\hat{B}'$ and
$\hat{C}$, $\hat{C}'$.
Since the operators $\hat{A}$, $\hat{B}$ and $\hat{C}$ operate in different Hilbert spaces, one can assume without loss of generality that all the vector products of type \eqref{eq.C3.2a} appearing in \eqref{eq.C3.1} are colinear with $\boldsymbol{e}_z$.
In this case all the unit vectors $\boldsymbol{a}$, $\boldsymbol{a}'$, $\boldsymbol{b}$, $\boldsymbol{b}'$, $\boldsymbol{c}$ and $\boldsymbol{c}'$ lie in the $xy$ plane.
Denoting as $\theta_a$ the angle between $\boldsymbol{a}'$ and $\boldsymbol{a}$, $\theta_b$ the angle between $\boldsymbol{b}'$ and $\boldsymbol{b}$, and $\theta_c$ the angle between $\boldsymbol{c}'$ and $\boldsymbol{c}$ (all these angles being in $[0,\pi]$), one gets
\begin{equation}\label{eq.C3.3}
\begin{split}
 \left(\hat{\mathscr{S}}{}^{(0|1|2)}\right)^2=&2+
2\cos\theta_a\cos\theta_b\cos\theta_c \\
& +2\sin\theta_a\sin\theta_b\, 
\hat{\Pi}_z^{(0)}\otimes\hat{\Pi}_z^{(1)}\otimes\mathds{1}^{(2)}\\
& +2\sin\theta_b\sin\theta_c\, 
\mathds{1}^{(0)}\otimes\hat{\Pi}_z^{(1)}\otimes\hat{\Pi}_z^{(2)}\\
& +2\sin\theta_a\sin\theta_c\, 
\hat{\Pi}_z^{(0)}\otimes\mathds{1}^{(1)}\otimes\hat{\Pi}_z^{(2)}.
\end{split}
\end{equation}
It is a simple matter to check that when $\theta_a$, $\theta_b$ and $\theta_c$ run through $[0,2\pi]^3$ the eigenvalues of the operator appearing in the right hand side of the above formula are all comprised in $[0,8]$. The extremal values 0 and 8 are reached for $(\theta_a,\theta_b,\theta_c)=(0,0,\pi)$ and $(\pi/2,\pi/2,\pi/2)$, respectively. It then follows that for any choice of the measurement directions
$\boldsymbol{a}$, $\boldsymbol{a}'$, $\boldsymbol{b}$, $\boldsymbol{b}'$, $\boldsymbol{c}$ and $\boldsymbol{c}'$
\begin{equation}\label{eq.C3.4}
\left\langle \hat{\mathscr{S}}{}^{(0|1|2)} \right\rangle^2 \le
\left\langle (\hat{\mathscr{S}}{}^{(0|1|2)})^2 \right\rangle \le 8,
\end{equation}
and thus the tripartite entanglement parameter \eqref{eq.TP7}
verifies
\begin{equation}\label{eq.C3.5}
S^{(0|1|2)}(\omega)\le 2\sqrt{2}.
\end{equation}
This is the tripartite equivalent of Cirel'son bound.

In the remaining of this Appendix we consider a related but somehow different problem: In order to violate as much as possible the Svetlichny inequality 
$S^{(0|1|2)}<2$ we aim at choosing 
measurement directions $\boldsymbol{a}$, $\boldsymbol{a}'$, $\boldsymbol{b}$, $\boldsymbol{b}'$, $\boldsymbol{c}$ and $\boldsymbol{c}'$ which 
maximize the expectation value of $\hat{\mathscr{S}}{}^{(0|1|2)}$. This can be done numerically as explained in Appendix \ref{app.maxi3}. We show here that this maximization can also be performed analytically in a particular instance.
Setting $\boldsymbol{a} \cdot\boldsymbol{a}' = \cos 2\theta $ with $\theta \in [ 0, \pi/2 ]$, we define two new unit vectors $\boldsymbol{a}_\pm$ by
\begin{equation} \label{eq.TP3}
 2\cos\theta \, \boldsymbol{a}_{+} = \boldsymbol{a} + \boldsymbol{a}',\;\;
2\sin\theta \, \boldsymbol{a}_{-} = \boldsymbol{a} - \boldsymbol{a}',
\end{equation}
where by definition $\boldsymbol{a}_{+}\cdot \boldsymbol{a}_{-} = 0$.
In terms of the new vectors \eqref{eq.TP3} the average of the three-mode operator \eqref{eq.TP0} reads
\begin{equation}
\langle \hat{\mathscr{S}}{}^{(0|1|2)} \rangle =
(W+Z) \cos\theta + (X-Y) \sin\theta,
\label{B_CHSH_analogue_3modes}
\end{equation}
where
\begin{equation} 
\begin{split}
W & =   \langle  \boldsymbol{a}_{+} \cdot \hat{\boldsymbol{\Pi}}{}^{(0)} \otimes 
\boldsymbol{b} \cdot \hat{\boldsymbol{\Pi}}{}^{(1)} \otimes 
\boldsymbol{c}'\cdot \hat{\boldsymbol\Pi}{}^{(2)} \rangle, \\
X & =   \langle \boldsymbol{a}_{-} \cdot \hat{\boldsymbol\Pi}{}^{(0)}  \otimes  \boldsymbol{b}' \cdot \hat{\boldsymbol\Pi}{}^{(1)}  \otimes \boldsymbol{c}' \cdot \hat{\boldsymbol\Pi}{}^{(2)}  \rangle, \\
Y & =   \langle  \boldsymbol{a}_{-} \cdot \hat{\boldsymbol\Pi}{}^{(0)}  \otimes \boldsymbol{b} \cdot \hat{\boldsymbol\Pi}{}^{(1)}  \otimes \boldsymbol{c} \cdot \hat{\boldsymbol\Pi}^{}{(2)} \rangle,\\
Z & = \langle  \boldsymbol{a}_{+} \cdot \hat{\boldsymbol\Pi}{}^{(0)}  \otimes \boldsymbol{b}' \cdot \hat{\boldsymbol\Pi}{}^{(1)}  \otimes \boldsymbol{c}\cdot  \hat{\boldsymbol\Pi}{}^{(2)} \rangle.
\end{split}
\label{WXYZ}
\end{equation}
It is convenient to introduce temporarily notations
\begin{equation}
W+Z={\cal A}\cos\delta,\quad X-Y={\cal A}\sin\delta,    
\end{equation}
which make it possible to cast \eqref{B_CHSH_analogue_3modes} under a simple form
\begin{equation}
 \langle \hat{\mathscr{S}}{}^{(0|1|2)} \rangle =    {\cal A} \cos (\theta - \delta).
\end{equation}
The maximum value of this expression is ${\cal A}$ and thus
\begin{equation} \label{eq.TP6}
\mbox{max}_{\theta}
\langle \hat{\mathscr{S}}{}^{(0|1|2)} \rangle =
 \sqrt{\left(W + Z \right)^2+\left(X-Y \right)^2}.
\end{equation}

The next step of the maximization procedure is easily performed at zero temperature and zero energy ($\omega=0$).
The reason is that, as illustrated in a typical case by \eqref{eq.NUM}, the explicit expressions of the $W$, $X$, $Y$, $Z$ coefficients in 
\eqref{WXYZ} and
\eqref{eq.TP6} involve combinations of terms of the type ${\cal T}_{rst}$ [as defined in Eq. \eqref{eq.W9}] 
that take on particularly simple values at $T=0$ and $\omega=0$.
Firstly, 
all the ${\cal T}_{rst}$ with at least one component along $x$ vanish. Therefore, the contribution of the components along $x$ of the different vectors involved in (\refeq{WXYZ}) cancels. It is thus enough to perform the maximization only considering vectors $\boldsymbol{a}_\pm$, $\boldsymbol{b}$, $\boldsymbol{b}'$, $\boldsymbol{c}$ and $\boldsymbol{c}'$ 
lying in the $y,z$ plane. 
A second simplification stems from the fact that, in the $y,z$ plane, all the ${\cal T}_{rst}$'s with an odd number of $y$ cancel at $T=0$ and $\omega=0$. In the waterfall and delta peak configurations\footnote{The situation is slightly different in the case of a flat profile configuration, where ${\cal T}_{yyz} = -1$ and ${\cal T}_{zzz} = {\cal T}_{zyy} = {\cal T}_{yzy} = 1$, with all other coefficients also vanishing. Once this modification is accounted for, the maximization procedure yields the same final result.} the only non-zero coefficients are (see the discussion in Appendix \ref{app.averages})
\begin{equation}
{\cal T}_{zzz} = {\cal T}_{yyz} = {\cal T}_{yzy} = - {\cal T}_{zyy} = 1.
\label{Tcoeffzeroener}
\end{equation}
In this case, the particular choice
\begin{equation}
\boldsymbol{a}_+ = \boldsymbol{b} = \boldsymbol{c}' =\boldsymbol{e}_z, \quad
\boldsymbol{a}_- = - \boldsymbol{b}' = - \boldsymbol{c} =\boldsymbol{e}_y,
\label{eq.param1}
 \end{equation}
plugged in Eqs. \eqref{WXYZ} leads to
\begin{equation}
\begin{split}
& W+Z={\cal T}_{zzz}+{\cal T}_{zyy}=2,\\        
& X-Y=-{\cal T}_{yyz}+{\cal T}_{yzy}=2.\\        
\end{split}
\end{equation}
and expression \eqref{eq.TP6} then shows that the upper bound \eqref{eq.C3.5} is reached. It is therefore not possible that another arrangement of vectors
 $\boldsymbol{a}$, $\boldsymbol{a}'$, $\boldsymbol{b}$, $\boldsymbol{b}'$, $\boldsymbol{c}$ and $\boldsymbol{c}'$
reaches a higher value and thus
\begin{equation}
S^{(0|1|2)}(\omega=0) \underset{T = 0}{=}  2\sqrt{2}.
\label{eq.B00max}
\end{equation}
The situation is completely different at finite temperature. In this case, as explained in Appendix \ref{app.averages} all the averages of the type
\eqref{eq.W9} cancel when $\omega\to 0$. It then follows that the quantities \eqref{WXYZ} behave in the same way and thus
\begin{equation}
S^{(0|1|2)}(\omega=0) \underset{T \ne 0}{=}  0.
\label{eq.B0T}
\end{equation}

\section{Results for other types of analog black holes}\label{app.other}

The results we have obtained have been presented in the main text for a waterfall configuration. The reason is that this is the only one which has been realized experimentally so far \cite{De_nova_2019,Kolobov_2021}. 
For completeness -- and also for emphasizing the typicality of the results presented in the main text --  we present in this appendix some equivalent results for the delta peak and flat profile configurations defined in Appendix \ref{app.different}. 

\begin{figure}
\centering
\includegraphics[width=\linewidth]{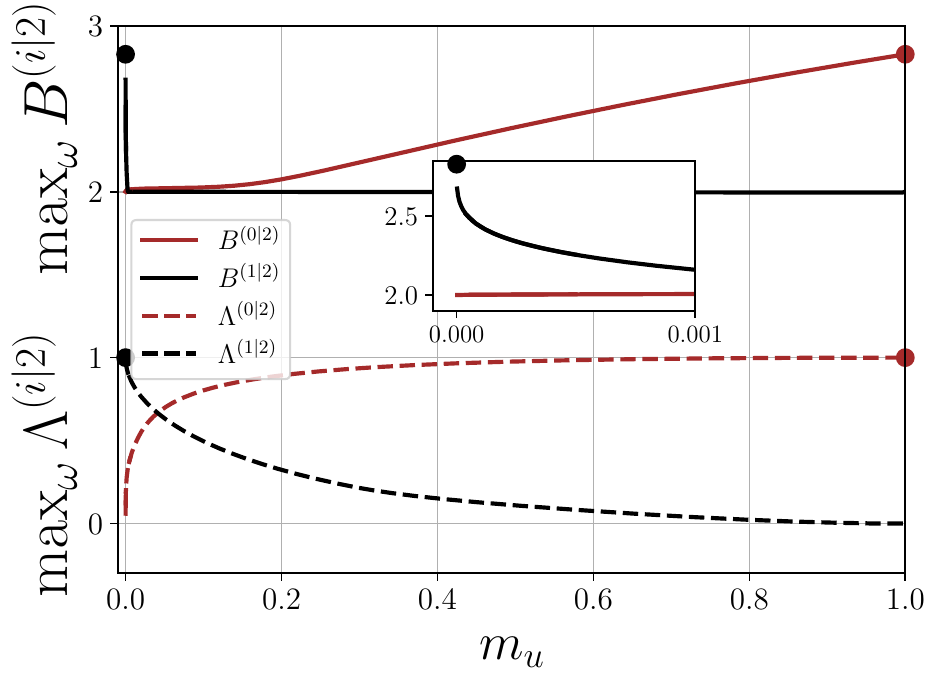}
\caption{Same as Fig. \ref{fig3:plus} for zero temperature delta peak
  configurations. The inset is a blow up at low $m_u$.}
\label{fig.H1}
\end{figure}
\begin{figure}
\centering
\includegraphics[width=\linewidth]{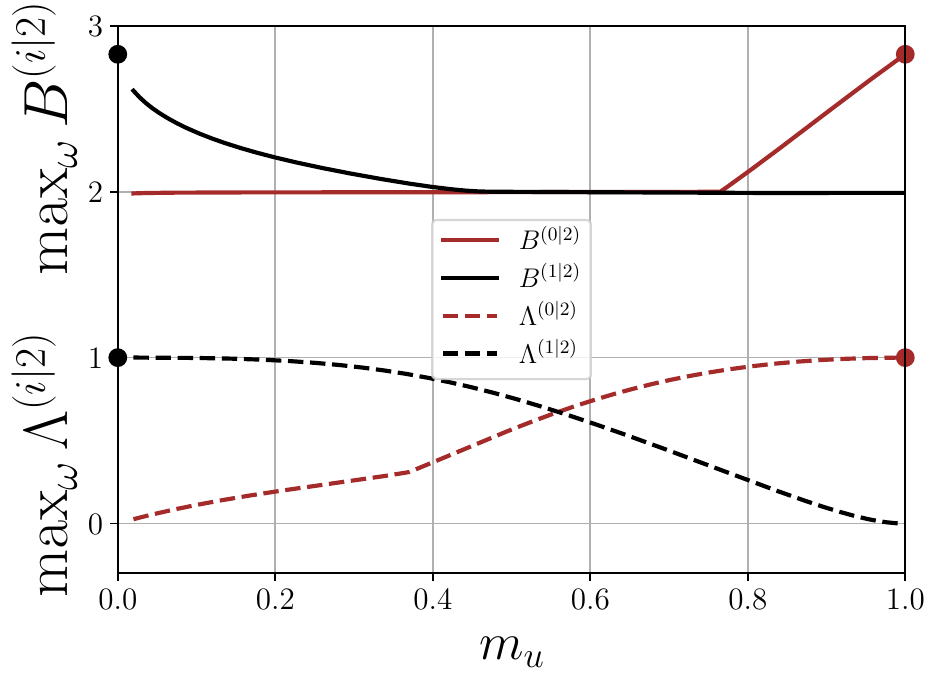}
\caption{Same as Fig. \ref{fig3:plus} for zero temperature flat
  profile configurations with $m_d=1/m_u^2$.}
\label{fig.H2}
\end{figure}
It is in particular of interest to plot the equivalents of
Fig. \ref{fig3:plus} for these alternative configurations. This is
done in Figs. \ref{fig.H1} and \ref{fig.H2}. In the flat profile
configuration the value of $m_d$ is a free parameter. Thus, for
comparing the results of the flat profile configuration with
Fig. \ref{fig3:plus} we impose $m_d=1/m_u^2$, as is the case for the
waterfall configuration, cf. Eq. \eqref{eq.model4a}. Such a procedure
is not required (nor possible) for the delta peak configuration where
fixing $m_u$ unambiguously determines $m_d$, see \eqref{eq.model4b}.
Fig. \ref{fig.H1} shows that the delta peak configuration has a
special feature: the EPR state formed by the companion and the partner
(see Sec. \ref{sec.Bell2}) is only reached at extremely low values of
the upstream mach number $m_u$, whereas bipartite nonseparability
between the Hawking quantum and the partner (modes 0 and 2,
respectively) is significant in a wide range of values of $m_u$
(roughly speaking, for $m_u \gtrsim 0.2$). Despite this peculiarity,
it is fair to say that Figs. \ref{fig.H1} and \ref{fig.H2} both
display the same general trend as Fig. \ref{fig3:plus}, supporting the
idea that the behavior discussed in the main text is generic. We do
not produce finite temperature figures equivalent to
Fig. \ref{fig3.pluT} in order not to overload the paper and because,
as expected, the delta peak and the flat profile configurations behave
similarly to the waterfall configuration when the temperature is
increased.  There is however a quantitative change which is worth
noticing: in these two alternative configurations the violation of
bipartite Bell inequality does not persist as much as for the
waterfall configuration when $T$ is increased.

It is also quite interesting to study genuine tripartite nonlocality in the delta peak and flat profile configurations by computing the zero temperature value of the Svetlichny parameter $S^{(0|1|2)}(\omega)$ and by then presenting
the equivalents of Fig. \ref{fig.Trip0}.
\begin{figure}
    \centering
    \includegraphics[width=\linewidth]{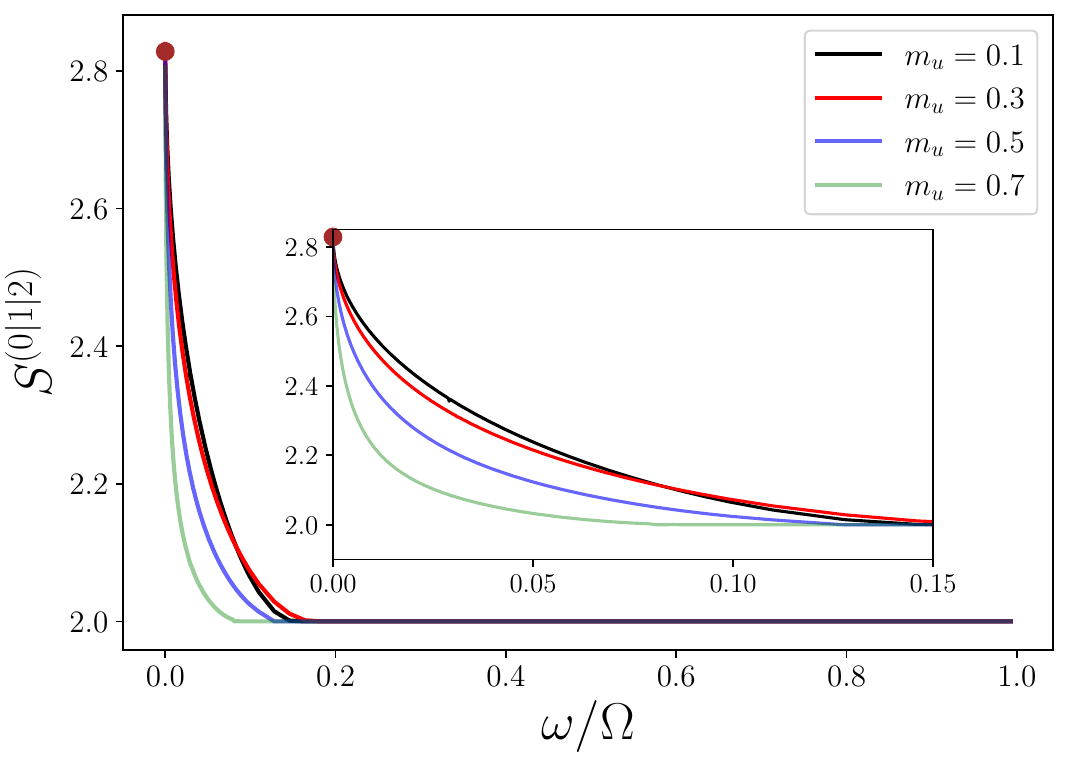}
    \caption{Same as Fig. \ref{fig.Trip0} for zero temperature delta
      peak configurations. The inset displays a blow up of the figure
      at low energy.}
    \label{fig.Trip0_DP}
\end{figure}
\begin{figure}
    \centering
    \includegraphics[width=\linewidth]{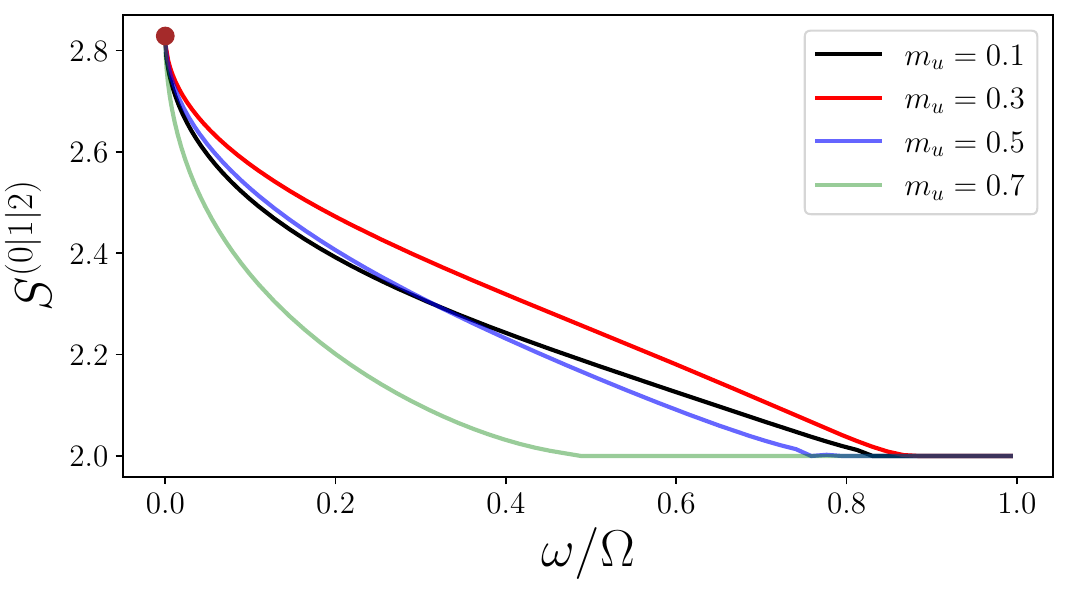}
    \caption{Same as Fig. \ref{fig.Trip0} for zero temperature flat
      profile configurations.}
    \label{fig.Trip0_FP}
\end{figure}
This is done in Figs. \ref{fig.Trip0_DP} and \ref{fig.Trip0_FP}. Here
again, the phenomenology is the same as the one discussed in the main
text: There is a clear signal of nonlocality at $T=0$, in a domain of
energy typically more extended than for the waterfall configuration,
but as is the case for the waterfall configuration this signal does
not persist at small finite temperature (the corresponding plots are
not shown in order not to overload the paper).

\begin{figure}
    \centering
    \includegraphics[width=\linewidth]{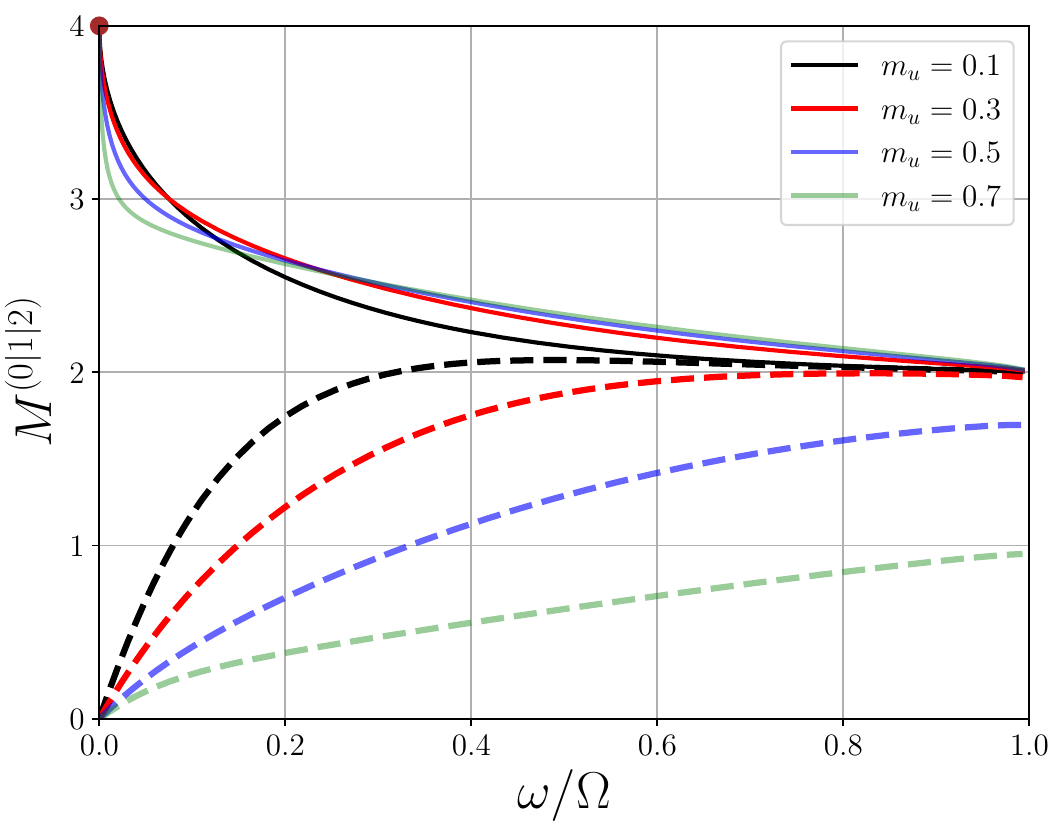}
    \caption{Same as Fig. \ref{fig.Mermin_WF} for delta peak
      configurations at $T=0$ (thin solid lines) and $T=0.1 g n_u$
      (thick dashed lines).}
    \label{fig.Mermin_DP}
\end{figure}
\begin{figure}
    \centering
    \includegraphics[width=\linewidth]{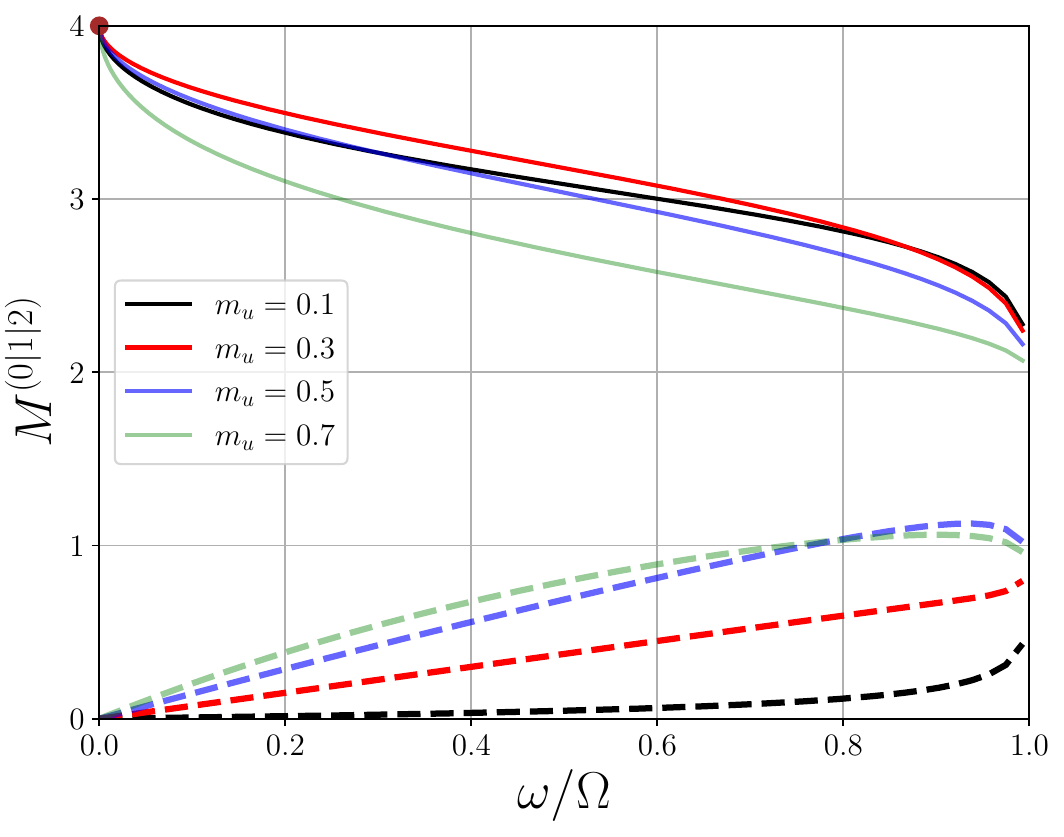}
    \caption{Same as Fig. \ref{fig.Mermin_WF} for flat profile
      configurations at $T=0$ (thin solid lines) and $T=0.1 g n_u$
      (thick dashed lines).}
    \label{fig.Mermin_FP}
\end{figure}
Finally, it is also important to address the GHZ character of the long
wave length modes in the delta peak and flat profile
configurations. As argued in Sec. \ref{sec.Bell3} of the main text,
both configurations display the GHZ paradox at $\omega=0$ and verify
$M^{(0|1|2)}(0)=4$ at zero temperature.  The behavior of the optimized
Mermin parameter $M^{(0|1|2)}(\omega)$ \eqref{eq.M1} is represented at
finite and zero temperature in Figs. \ref{fig.Mermin_DP} and
\ref{fig.Mermin_FP} which correspond to the delta peak and flat
profile configuration, respectively. At zero temperature the signal
\eqref{eq.M3a} of genuine tripartite entanglement is more pronounced
for the delta peak and flat profile configurations that for the
waterfall. This was also the case for the signal \eqref{eq.M3b} of
genuine tripartite nonlocality.  However, this signal, although more
noticeable at $T=0$, is less resilient to an increase of temperature
than for the waterfall configuration. The deleterious effect of
temperature is particularly pronounced for the flat profile
configuration, see Fig. \ref{fig.Mermin_FP}. However, although it is
clear that each configuration bares its specificities, it is also
clear than the general trend is the same: the departure from the GHZ
signal $M^{(0|1|2)}(0)=4$ increases at high energy. At finite
temperature, the GHZ behavior is lost. The Mermin parameter
nonetheless provides a signal of nonlocality if larger than 2. For
that matter, the signal is more pronounced at $T\neq 0$ for the
waterfall configuration than for the two others.

\newpage

\bibliography{biblio}

\begin{thebibliography}{116}
\bibitem [{rem()}]{rem_local}%
  \BibitemOpen
  \href@noop {} {}\bibinfo {note} {From local realism is follows that the
  measurement of an observable of the type
  $\boldsymbol{a}\cdot\hat{\boldsymbol{\Pi}}{}^{(i)} \otimes
  \boldsymbol{b}\cdot\hat{\boldsymbol{\Pi}}{}^{(j)}$ is the product
  $\Pi^{(i)}(\lambda,\boldsymbol{a})\cdot \Pi^{(j)}(\lambda,\boldsymbol{b})$ of
  the result $\Pi^{(i)}(\lambda,\boldsymbol{a})=\pm 1$ of measurement of the
  pseudo-spin $\hat{\boldsymbol{\Pi}}{}^{(i)}$ along direction $\boldsymbol{a}$
  with the result $\Pi^{(i)}(\lambda,\boldsymbol{b})=\pm 1$ of measurement of
  the pseudo-spin $\hat{\boldsymbol{\Pi}}{}^{(j)}$ along direction
  $\boldsymbol{b}$, where $\lambda$ is the hidden variable. Within this
  approach, there are only two possible cases: either
  $\Pi^{(i)}(\lambda,\boldsymbol{a})$ and $\Pi^{(i)}(\lambda,\boldsymbol{a}')$
  are equal (their difference cancels and they sum to $\pm 2$) either they are
  opposite (their sum cancels and their difference is $\pm 2$). From expression
  \eqref{eq.GKMR3} directly follows that such a measurement of
  $\hat{\mathscr{B}}{}^{(i|j)}$ yields a result $\pm 2$.}\BibitemShut {Stop}%
\end{thebibliography}

\end{document}